\documentclass[a4paper]{article}
\usepackage{amssymb,amsmath,mathrsfs}
\usepackage{graphicx}

\usepackage{amsthm}
\newenvironment{acknowledgement}{\section*{Acknowledgments}}{}
\newtheorem{theorem}{Theorem}
\newtheorem{proposition}[theorem]{Proposition}
\newtheorem{definition}[theorem]{Definition}
\newtheorem{lemma}[theorem]{Lemma}
\newtheorem{corollary}[theorem]{Corollary}

\providecommand{\abs}[1]{{\lvert{#1}\rvert}}
\providecommand{\bra}[1]{{\langle{#1}\rvert}}
\providecommand{\ket}[1]{{\lvert{#1}\rangle}}
\providecommand{\bracket}[2]{{\langle{#1}|{#2}\rangle}}
\providecommand{\openone}{{1\!\!1}}
\newcommand{\ketbra}[2]{{\ket{#1}\!\bra{#2}}}

\DeclareMathOperator{\vspan}{span}
\DeclareMathOperator{\rank}{rank}
\DeclareMathOperator{\tr}{tr}
\DeclareMathOperator{\supp}{supp}

\DeclareMathOperator{\Real}{Re}

\providecommand{\etal}{\emph{et al.}}
\newcommand{\mx}[1]{\quad \text{#1} \quad}
\newcommand{\apriori}{\emph{a priori}}
\newcommand{\e}[1]{\mathrm e^{#1}}
\newcommand{\pr}[1]{\ketbra{#1}{#1}}
\newcommand{\id}{\openone}
\newcommand{\Hilbert}{{\mathscr H}}

\renewcommand{\Re}{\Real}

\begin{document}
\title{Structural approach to unambiguous discrimination of two mixed quantum 
states}
\author{M.~Kleinmann$^{1,2}$\thanks{electronic address: 
                                    \texttt{Matthias.Kleinmann@uibk.ac.at}},
        H.~Kampermann$^1$, and
        D.~Bru\ss$^1$}
\date{}

\maketitle
\begin{center}\em
 $^1$Institut f\"ur Theoretische Physik III,
 Heinrich-Heine-Universit\"at D\"usseldorf,
 D-40225 D\"usseldorf,
 Germany

 $^2$Institut f\"ur Quantenoptik und Quanteninformation,
 \"Osterreichische Akademie der Wissenschaften,
 A-6020 Innsbruck,
 Austria
\end{center}

\begin{abstract}
We analyze the optimal unambiguous discrimination of two arbitrary mixed 
 quantum states.
We show that the optimal measurement is unique and we present this optimal 
 measurement for the case where the rank of the density operator of one of the 
 states is at most 2 (``solution in 4 dimensions'').
The solution is illustrated by some examples.
The optimality conditions proved by Eldar \etal\ [Phys.\ Rev.\ A {\bf 69}, 
 062318 (2004)] are simplified to an operational form.
As an application we present optimality conditions for the measurement, when 
 only one of the two states is detected.
The current status of optimal unambiguous state discrimination is summarized 
 via a general strategy.
\end{abstract}

\thispagestyle{empty}
\tableofcontents

% vim:ft=tex ==================================================================%
\section{Introduction}
Among the subtleties in quantum information processing and in quantum 
 communication protocols are the properties that originate from the fact that 
 in quantum mechanics non-orthogonal states cannot be discriminated perfectly.
In the most na\"{i}ve approach to quantum state discrimination -- the minimum 
 error discrimination (cf.\ Ref.~\cite{Helstrom:1976,Herzog:2004PRA}) -- this 
 leads to the fact, that the identification of a state might be erroneous with 
 some finite probability.
Ivanovic \cite{Ivanovic:1987PLA} and Dieks \cite{Dieks:1988PLA} showed that one 
 can avoid erroneous measurement results and that a measurement with a 
 conclusive state identification is possible.
In the case of non-orthogonal states, this strategy cannot work with a success 
 probability of one.
Peres showed in Ref.~\cite{Peres:1988PLA} how the optimum of this success 
 probability can be achieved in the case of pure states, both having the same 
 \apriori\ probability.
The discussion of the optimal unambiguous discrimination of two pure states was 
 completed by Jaeger and Shimony in Ref.~\cite{Jaeger:1995PLA}.
They derived the optimal solution for arbitrary \apriori\ probabilities.

Although it was long ago stated to be an interesting problem 
 \cite{Bennett:1996PRA}, the unambiguous discrimination of \emph{mixed} states 
 did not attract much attention for a long time.
This changed with an example introduced by Sun \etal\ in 
 Ref.~\cite{Sun:2002PRA} and the first general analysis of the unambiguous 
 discrimination of mixed states by Rudolph \etal\ in 
 Ref.~\cite{Rudolph:2003PRA}.
After that, several general results and special classes of optimal solutions 
 were found, cf.\ 
 Ref.~\cite{Bergou:2003PRL,Raynal:2003PRA,Bergou:2005PRA,Herzog:2005PRA,%
 Raynal:2005PRA,Herzog:2007PRA,Raynal:2007PRA}.
While Bergou \etal\ derived in Ref.~\cite{Bergou:2003PRL,Bergou:2005PRA} the 
 optimal measurement for the unambiguous discrimination of a pure state and an 
 arbitrary mixed state, no analysis so far did succeed to produce a general 
 solution for the simplest instance of genuine mixed state discrimination, the 
 discrimination of two mixed states where both density operators have a rank of 
 2.
Also the simple question whether the optimal measurement in general is unique 
 remained unanswered.

The answers to these two questions are among the central results of this 
 contribution.
The uniqueness of the optimal measurement is stated in Proposition~\ref{p14866} 
 and the general solution for rank 2 density operators is presented in 
 Sec.~\ref{s17133}.

A valuable tool to approach both questions turned out to be a result by Eldar 
 \etal\ in Ref.~\cite{Eldar:2004PRA}.
They showed necessary and sufficient conditions for a given measurement to be 
 optimal.
However, these conditions are difficult to verify, since the criterion implies 
 the proof of the existence or non-existence of an operator with certain 
 properties.
In Corollary~\ref{c24797} we reformulate this criterion in such a way, that it 
 can be directly applied to a given measurement.
As a further immediate consequence of this Corollary we will be able to provide 
 simple optimality conditions for a very special type of measurement:
The measurement which only detects one out of the two states, cf.\ 
 Sec.~\ref{s194}.
It will also become possible to provide a simple proof and a deeper insight 
 into the fidelity form measurement \cite{Herzog:2005PRA,Raynal:2005PRA}, cf.\ 
 Sec.~\ref{s3330}.

Before we arrive at these results, we first provide an analysis of unambiguous 
 state discrimination (USD), beginning in Sec.~\ref{s31965}, where we derive 
 general results and continuing in Sec.~\ref{s5936}, in which we specialize 
 to the optimal case.

An analysis of the structure of the optimal measurement in particular yields 
 Theorem~\ref{t4134}.
This Theorem is a cornerstone in order to prove the uniqueness of the optimal 
 measurement and also provides a simple proof of the ``second reduction'' shown 
 by Raynal \etal\ in Ref.~\cite{Raynal:2003PRA}.
We summarize and deepen the analysis carried out in Ref.~\cite{Raynal:2003PRA} 
 in Proposition~\ref{p6437}, Proposition~\ref{p2770}, and Lemma~\ref{l2980}.

In Sec.~\ref{s23733} we will provide a generic scheme in order to approach a 
 given optimization problem for USD.
We conclude in Sec.~\ref{s23013}.

% vim:ft=tex ==================================================================%
\section{Defining properties of USD}\label{s31965}
%------------------------------------------------------------------------------%
\subsection{Main definitions}
In quantum state discrimination of $n$ quantum states it is usually assumed 
 that the density operators $\rho_1, \dotsc, \rho_n$ of all possible input 
 states are known, together with the probability $p_1, \dotsc, p_n$ of their 
 occurrence.
For $1\le \mu\le n$, the \apriori\ probability $p_\mu\ge 0$ and the 
 corresponding density operator $\rho_\mu \ge 0$ with $\tr(\rho_\mu)= 1$ 
 naturally combine to a \emph{weighted density operator} $\gamma_\mu= 
 p_\mu\rho_\mu$.
Hence the trace of a weighted density operator $\gamma_\mu$ is the \apriori\ 
 probability of the state, $\tr(\gamma_\mu)= p_\mu$.
Using this notation, the input states are represented by a family of positive 
 semi-definite operators $\mathcal S= (\gamma_\mu)$.
For a meaningful interpretation in terms of probability, we clearly need to 
 have $\sum_\mu \tr(\gamma_\mu)= 1$.
However, we will not require this normalization, as the subsequent definition 
 and analysis is independent of it, and for certain statements (cf.\ e.g.\ 
 Proposition~\ref{p18391}) it will be useful to explicitly allow $\sum_\mu 
 \tr(\gamma_\mu) < 1$.

In the following we will only consider the case of two input states, i.e., 
 $\mu=1,2$.
We restrict our analysis to finite-dimensional quantum systems, such that any 
 possible quantum state of the system can be represented by a density operator 
 which acts on a Hilbert space $\Hilbert$ of finite dimension.
We will use the formalism of generalized measurements in which a physical 
 measurement with $M$ possible outcomes is described by a positive operator 
 valued measure $\mathcal E= (E_1,\dotsc,E_M)$ on $\Hilbert$, i.e., by a family 
 of $M$ positive semi-definite operators which sum up to the identity, $\sum_k 
 E_k= \id$.

Let us introduce our notation.
We denote by $\ker A= \{ \ket k\in\Hilbert \mid A\ket k= 0\}$ the kernel of an 
 operator $A$, and we write $A\Hilbert= \{ A \ket\phi \mid \ket\phi \in 
 \Hilbert \}$ for its image.
The support of a positive semi-definite operator $\rho$ is written as $\supp 
 \rho= \{ \ket\phi\in \Hilbert \mid \exists\, \alpha>0 \colon \rho- \alpha \pr 
 \phi \ge 0\}$.
Note, that the support of $\rho$ is the orthocomplement of its kernel, 
 $\supp\rho= (\ker\rho)^\perp$ and since $\rho$ is self-adjoint, $\rho\Hilbert= 
 \supp\rho$ holds.

By a \emph{projector} we always mean an orthogonal projector, unless we 
 explicitly state that the projector is oblique (cf.\ Lemma~\ref{l10547} in 
 Appendix~\ref{s26828}).
We use upper case Greek letters for orthogonal projectors,
 $\Sigma^\dag= \Sigma= \Sigma^2$.
The symbols ``$\subset$'' and ``$\supset$'' are used such that they also 
 include equality, i.e., $\mathscr A= \mathscr B$ if and only if $\mathscr A 
 \subset \mathscr B$ and $\mathscr A \supset \mathscr B$.

For a pair of weighted density operators $\mathcal S= (\gamma_1, \gamma_2)$, we 
 abbreviate
\begin{equation}
 \supp\mathcal S \equiv \supp(\gamma_1+ \gamma_2)
                 =\supp\gamma_1+ \supp\gamma_2,
\end{equation}
 for the collective support of $\mathcal S$, which is the physically relevant 
 subspace for the discrimination task and $\ker\mathcal S$ for the common 
 kernel of $\mathcal S$, which then is the trivial subspace,
\begin{equation}
 \ker\mathcal S \equiv \ker(\gamma_1+ \gamma_2)
                =\ker\gamma_1\cap \ker\gamma_2.
\end{equation}
The task of optimal unambiguous discrimination of two mixed states is defined 
 as follows.
\begin{definition}
A positive operator valued measure $\mathcal E=(E_1,E_2,E_?)$ is called an 
 \emph{unambiguous state discrimination (USD) measurement} of a pair of 
 weighted density operators $\mathcal S=(\gamma_1,\gamma_2)$ if 
 $\tr(E_2\gamma_1)= 0$ and $\tr(E_1\gamma_2)= 0$.
The \emph{success probability} $P_\mathrm{succ}$ of $\mathcal E$ of $\mathcal 
 S$ is given by
\begin{equation}
 P_\mathrm{succ}(\mathcal E; \mathcal S)= \tr(E_1\gamma_1)+ \tr(E_2\gamma_2).
\end{equation}

A USD measurement $\mathcal E$ of $\mathcal S$ is {\rm optimal} if it has 
 maximal success probability, i.e., if for any USD measurement $\mathcal E'$ of 
 $\mathcal S$, $P_\mathrm{succ}(\mathcal E;\mathcal S)\ge 
 P_\mathrm{succ}(\mathcal E';\mathcal S)$ holds.
A USD measurement $\mathcal E$ of $\mathcal S$ is called {\rm proper} if
 $\supp(E_1+ E_2)\subset \supp\mathcal S$.
\end{definition}

The condition $\tr(E_2\gamma_1)= 0$ is equivalent to $\supp E_2\subset 
 \ker\gamma_1$ and $\tr(E_1\gamma_2)= 0$ is equivalent to $\supp E_1\subset 
 \ker\gamma_2$.
Thus it is simple to write down some USD measurement for a given pair $\mathcal 
 S$.
In the next section we will see, that it is sufficient to consider proper USD 
 measurements.
But the set of proper USD measurements in particular is compact (this follows 
 from the above definition or more directly from Proposition~\ref{p6437}) and 
 hence there always exists at least one proper USD measurement, which maximizes 
 the success probability.

%------------------------------------------------------------------------------%
\subsection{Trivial subspaces}\label{s20689}
For any USD measurement $\mathcal E=(E_1,E_2,E_?)$ of $\mathcal S=(\gamma_1, 
 \gamma_2)$ one readily constructs a proper USD measurement $\mathcal 
 E'=(E_1',E_2',E_?')$ with the same marginal probabilities, i.e., 
 $\tr(E_1\gamma_1)= \tr(E_1'\gamma_1)$ and $\tr(E_2\gamma_2)= 
 \tr(E_2'\gamma_2)$.
For that the most straightforward approach is to choose $E_1'$ and $E_2'$ to be 
 the projection of $E_1$ and $E_2$ onto $\supp\mathcal S$ and to set $E_?'=\id- 
 E_1'- E_2'$.

As an important feature of proper USD measurements we will show that the 
 optimal proper USD measurement is unique (cf.\ Proposition~\ref{p14866}).
Such a statement of uniqueness clearly can only hold if we require that the 
 measurement is proper.
For illustrative reasons let us provide an example of an optimal USD 
 measurement, which is not proper and where the measurement operators do not 
 even commute with the projector onto $\supp\mathcal S$:
We consider two non-orthogonal pure states with
\begin{equation}
 \gamma_1=\tfrac12 \pr1,\quad
 \gamma_2=\tfrac12 \pr+,
\end{equation}
 where $\ket+= (\ket0+ \ket1)/\sqrt2$ and the measurement $\mathcal E=(E_1, 
 E_2, \id- E_1- E_2)$ with
\begin{equation}
 E_\mu =(3-3/\sqrt2) \pr{e_\mu},
\end{equation}
 where $\ket{e_1}= (\ket0- \ket1- \ket2)/\sqrt3$ and $\ket{e_2}= (\sqrt2\ket 0+ 
 \ket 2)/\sqrt3$.
It is straightforward to verify, that this measurement is a USD measurement and 
 has a success probability of $P_\mathrm{succ}=1-1/\sqrt2$ as given by the 
 optimal solution due to Peres \cite{Peres:1988PLA}.

The subspace $\ker\mathcal S$ cannot play any role in USD, since the support of 
 $\gamma_1$ and $\gamma_2$ is orthogonal to this space.
Similarly, the subspace $\supp\gamma_1\cap \supp\gamma_2$ necessarily is 
 orthogonal to the support of $E_1$ and $E_2$, since $\supp E_1\subset 
 \ker\gamma_2$ and $\supp E_2\subset \ker\gamma_1$.
The following proposition is a consequence of this observation:
\begin{proposition}[cf.\ Theorem~1 in Ref.~\cite{Raynal:2003PRA}]\label{p18391}
Let $\mathcal S= (\gamma_1, \gamma_2)$ be a pair of weighted density operators.
Denote by $\Pi_\nparallel$ the projector onto $(\ker\gamma_1+ \ker\gamma_2)$ 
 and write $\mathcal S^\nparallel= (\Pi_\nparallel\gamma_1\Pi_\nparallel, 
 \Pi_\nparallel\gamma_2\Pi_\nparallel)$ for the projected pair.
Let $q\ge 0$.

Then $\mathcal E$ is a proper USD measurement for $\mathcal S$ with 
 $P_\mathrm{succ}(\mathcal E; \mathcal S)= q$ if and only if $\mathcal E$ is a 
 proper USD measurement for $\mathcal S^\nparallel$ with 
 $P_\mathrm{succ}(\mathcal E; \mathcal S^\nparallel)= q$.
\end{proposition}
\noindent (Note, that $(\ker \gamma_1+ \ker\gamma_2)$ is the orthocomplement of 
 $(\supp\gamma_1\cap \supp\gamma_2)$, which can be considered to be the 
 ``parallel'' part of the support of $\gamma_1$ and $\gamma_2$.)
\begin{proof}
If $\mathcal E$ is a USD measurement of $\mathcal S$, we have $\Pi_\nparallel 
 E_\mu \Pi_\nparallel= E_\mu$ and so clearly $\tr(E_\mu \gamma_\nu)= \tr(E_\mu 
 \Pi_\nparallel \gamma_\nu \Pi_\nparallel)$ holds.
We have that $\supp E_\mu\subset \supp\mathcal S$ and $\supp E_\mu\subset 
 (\ker\gamma_1+ \ker\gamma_2)$.
Due to $\supp\mathcal S^\nparallel= \supp\mathcal S \cap (\ker\gamma_1+ 
 \ker\gamma_2)$, it follows that $\mathcal E$ is also proper for $\mathcal 
 S^\nparallel$.

For the converse, since $\mathcal E$ is proper for $\mathcal S^\nparallel$ we 
 have in particular $\Pi_\nparallel E_\mu \Pi_\nparallel= E_\mu$ and hence 
 $\tr(E_\mu \Pi_\nparallel \gamma_\nu \Pi_\nparallel)= \tr(E_\mu \gamma_\nu)$.
Furthermore we have $\supp E_\mu \subset \supp\mathcal S^\nparallel= 
 \supp\mathcal S \cap (\ker\gamma_1+ \ker\gamma_2)$, i.e., $\mathcal E$ is 
 proper for $\mathcal S$.
\end{proof}

%------------------------------------------------------------------------------%
\subsection{The role of $E_?$}
For the discussion of USD measurements it is useful to note that the 
 measurement operator corresponding to the inconclusive result, $E_?$, already 
 completely determines a proper USD measurement.
\begin{proposition}\label{p6437}
For an operator $E_?$ and a pair of weighted density operators $\mathcal 
 S=(\gamma_1, \gamma_2)$ there exist operators $E_1$ and $E_2$, such that 
 $\mathcal E= (E_1, E_2, E_?)$ is a proper USD measurement of $\mathcal S$, if 
 and only if $E_?$ acts as identity on $\ker\mathcal S$, $E_?\ge 0$, $\id- 
 E_?\ge 0$ and $\gamma_1 (\id - E_?) \gamma_2= 0$.

Given $E_?$, the proper USD measurement $\mathcal E$ of $\mathcal S$ is unique.
\end{proposition}
\begin{proof}
It is straightforward to see that the conditions are necessary.
The proof of sufficiency and uniqueness is constructive:
Let us write $Q_1$ for the oblique projector from $\ker\gamma_2\cap 
 \supp\mathcal S$ to $\supp\gamma_1\cap (\ker\gamma_1+ \ker\gamma_2)$ (for a 
 brief introduction to oblique projectors cf.\ Lemma~\ref{l10547} in 
 Appendix~\ref{s26828}).
Then we have for any proper USD measurement $E_2 Q_1= 0$ and $E_1= E_1 Q_1$.
Hence
\begin{equation}
 E_1= Q_1^\dag (E_1 + E_2) Q_1= Q_1^\dag (\id- E_?) Q_1
\end{equation}
 is the only candidate for $E_1$, given $E_?$.
Due to $\id- E_?\ge 0$, this construction ensures that $E_1 \ge 0$.
An analogous construction holds for $E_2$.

It remains to show that $E_1+ E_2- (\id -E_?)= 0$.
We decompose the Hilbert space into the sum
\begin{equation}\begin{split}
 \Hilbert&= \ker\mathcal S    \oplus (\supp\gamma_1+ \supp\gamma_2).
\end{split}\end{equation}
With $\Pi_\perp$ the projector onto $\ker\mathcal S$, we have $E_\mu \Pi_\perp= 
 0$ and since $E_?$ acts as identity on $\ker\mathcal S$, also $(\id- 
 E_?)\Pi_\perp= 0$ holds.
Using, that by construction $\gamma_1 E_\mu \gamma_2= 0$, we furthermore have
\begin{equation}
 \gamma_1[E_1+ E_2- (\id- E_?)]\gamma_2= -\gamma_1(\id- E_?)\gamma_2= 0.
\end{equation}

From $\gamma_1(\id - E_?)\gamma_2= 0$ and $\id- E_?\ge 0$ it follows that with 
 $\Pi_\parallel$ the projector onto $\supp\gamma_1\cap \supp\gamma_2$, we have 
 $(\id- E_?)\Pi_\parallel= 0$.
Furthermore, one verifies that $Q_1 \gamma_1= (\id-\Pi_\parallel)\gamma_1$ and 
 hence $\gamma_1(E_1+ E_2) \gamma_1= \gamma_1 E_1 \gamma_1= \gamma_1(\id- 
 E_?)\gamma_1$.
A similar argument for $\gamma_2$ finishes the proof.
\end{proof}
Due to this Proposition~\ref{p6437} we sometimes refer to an operator $E_?$ as 
 a proper USD measurement if it satisfies the conditions of the Proposition.
In addition from the Proposition it follows easily that the set of USD 
 measurements is bounded and closed and hence in particular it is compact.

A proper USD measurement is already uniquely defined by $E_?(\gamma_2- 
 \gamma_1)E_?$ (as it will turn out below, cf.\ Lemma~\ref{l17322}, in the 
 optimal case this operator is in some sense much simpler than $E_?$ itself).
Namely, with $\Pi_\perp$ the projector onto $\ker\mathcal S$ and $(\gamma_1+ 
 \gamma_2)^-$ denoting the inverse of $(\gamma_1+ \gamma_2)$ on its support we 
 have the identity
\begin{equation}\label{e5125}
\begin{split}
E_?= \Pi_\perp+ (\gamma_1+ \gamma_2)^-&\big\{\gamma_1\gamma_2+ \gamma_2\gamma_1
     \\&\;+
   \sqrt{\gamma_1}
   \sqrt{\sqrt{\gamma_1}[\gamma_2-
       E_?(\gamma_2-\gamma_1) E_?]\sqrt{\gamma_1}}
   \sqrt{\gamma_1}
     \\&\;+
   \sqrt{\gamma_2}
   \sqrt{\sqrt{\gamma_2}[\gamma_1-
       E_?(\gamma_1-\gamma_2) E_?]\sqrt{\gamma_2}}
   \sqrt{\gamma_2}
   \\&\big\}(\gamma_1+ \gamma_2)^-.
\end{split}
\end{equation}
In order to see this, first note that using $\sqrt{\gamma_1}(\id- 
 E_?)\sqrt{\gamma_2}= 0$ and $E_? \ge 0$ the term in curly brackets can be 
 rewritten as
\begin{equation}
 (\gamma_1+ \gamma_2)^2- \gamma_1^2- \gamma_2^2+ 
 \sqrt{\gamma_1}\sqrt{(\sqrt{\gamma_1} E_? \sqrt{\gamma_1})^2} \sqrt{\gamma_1}
 + \gamma_2 E_?  \gamma_2.
\end{equation}
Then due to $(\gamma_1+ \gamma_2)^- (\gamma_1+ \gamma_2)= \id- \Pi_\perp$ and 
 once more $\gamma_1(\id- E_?)\gamma_2= 0$ we see that the right hand side of 
 Eq.~\eqref{e5125} is given by
\begin{multline}
 \Pi_\perp+(\gamma_1+ \gamma_2)^-(\gamma_1+ \gamma_2)[\id- (\id- 
 E_?)](\gamma_1+ \gamma_2) (\gamma_1+ \gamma_2)^-\\
   =  \Pi_\perp+ (\id-\Pi_\perp)E_?(\id-\Pi_\perp)
\end{multline}
This expression is equal to $E_?$, since for a proper measurement 
 $E_?\Pi_\perp= \Pi_\perp$ holds.

Using the forthcoming Lemma~\ref{l17322}, Eq.~\eqref{e5125}, and 
 Proposition~\ref{p6437}, it will become possible to reconstruct the optimal 
 measurement given only the projective part of $E_?$.
This projective part is given by $\ker(\id- E_?)$.
It has a very specific structure, which originates in the condition 
 $\gamma_1(\id - E_?)\gamma_2= 0$.
Let $\Pi_\mu$ denote the projector onto $\supp\gamma_\mu$ and $\Pi_\perp$ 
 denote the projector onto $\ker \mathcal S$.
For any proper measurement these projectors satisfy $\Pi_1 (\id- E_?) \Pi_2= 0$ 
 and $(\id- E_?)\Pi_\perp= 0$, and hence Lemma~\ref{l2502} 
 (Appendix~\ref{s26828}) applies, i.e., for any proper measurement,
\begin{multline}\label{e25118}
 \ker(\id- E_?)= \big(\{\ker(\id- E_?)\cap \supp\gamma_1\}\\
                 +\{\ker(\id- E_?)\cap \supp\gamma_2\}\big)
                 \oplus\ker \mathcal S
\end{multline}
 holds.
(Note, that in the right hand side of Eq.~\eqref{e25118} the first and second 
 term are in general not orthogonal and share $\supp\gamma_1\cap \supp\gamma_2$ 
 as a common subspace.)
Although this result may seem to be quite technical, in certain situations it 
 turns out to be a quite powerful tool.

% vim:ft=tex ==================================================================%
\section{Simple properties of optimal measurements}\label{s5936}
The following theorem makes a simple but fundamental statement about the 
 structure of optimal measurements.
It states that apart from trivial cases no vector which is in the kernel of 
 $\gamma_1$ or in the kernel of $\gamma_2$ will be in the support of $E_?$.
This clearly gives an upper bound on the rank of $E_?$.
On the other hand the condition $\gamma_1(\id - E_?)\gamma_2=0$ provides a 
 lower bound on the rank of $E_?$.
The second part of the theorem states that these bounds coincide and fix the 
 rank of $E_?$.
\begin{theorem}\label{t4134}
Let $\mathcal E=(E_1,E_2,E_?)$ be an optimal USD measurement for a pair of 
 weighted density operators $\mathcal S=(\gamma_1, \gamma_2)$.
Then $(\supp E_? \cap \ker \gamma_1)= (\supp E_? \cap \ker \gamma_2)$.

If $\mathcal E$ in addition is proper, then $\supp E_?\cap \ker\gamma_1= 
 \ker\mathcal S$, $\supp E_?\cap \ker\gamma_2=\ker\mathcal S$, and $\rank E_?= 
 \rank \gamma_1\gamma_2+ \dim\ker\mathcal S$.
\end{theorem}
\noindent
(Remember, that the rank of an operator $A$ is given by $\dim(A\Hilbert)\equiv 
 \dim \Hilbert- \dim \ker A$, i.e., the number of strictly positive eigenvalues 
 of $A^\dag A$.)
\begin{proof}
Let $\ket\phi\in \supp E_? \cap \ker \gamma_1$.
Then due to $\ket \phi\in \supp E_?$ there exists an $\alpha>0$ such that $E_?- 
 \alpha\pr \phi \ge 0$.
We define a new USD measurement by $\mathcal E'= (E_1, E_2+ \alpha\pr\phi, E_?- 
 \alpha\pr\phi)$.
From the optimality condition for $\mathcal E$, i.e., $P_\mathrm{succ}(\mathcal 
 E',\mathcal S)\le P_\mathrm{succ}(\mathcal E,\mathcal S)$, we find 
 $\alpha\bra\phi\gamma_2\ket\phi\le 0$ which only can hold if 
 $\gamma_2\ket\phi= 0$.
Since $\ket \phi\in \supp E_?$, $(\supp E_? \cap \ker \gamma_1)\subset (\supp 
 E_?  \cap \ker \gamma_2)$ follows.
An analogous argument holds for the ``$\supset$'' part and finishes the proof 
 of the first assertion.

From this result by intersection with $(\ker\gamma_1)$ one immediately finds 
 $(\supp E_?\cap \ker\gamma_1)= (\supp E_?\cap \ker\mathcal S)$.
In the case of a proper measurement, however, $\supp E_?\supset \ker\mathcal S$ 
 and hence $(\supp E_?\cap \ker\gamma_1)= \ker\mathcal S$ follows.

Let $E_?'$ denote $E_?$ projected onto $\supp\mathcal S$.
Since the measurement is proper, $E_?- E_?'$ is the projector onto 
 $\ker\mathcal S$ and $\supp E_?= \supp E_?'\oplus \ker\mathcal S$.
From the previous results we have $E_?'\Hilbert \cap \ker\gamma_2= \{0\}$ and 
 $E_?'\gamma_2\Hilbert \cap \ker\gamma_1= \{0\}$.
Then due to Lemma~\ref{l12912} (Appendix~\ref{s26828}) if follows 
 $\ker(\gamma_2 E_?')= \ker E_?'$ and $\ker (\gamma_1 E_?' \gamma_2)= \ker 
 (E_?' \gamma_2)$.
Hence,
\begin{equation}\begin{split}
  \dim\ker E_?'&= \dim\ker(\gamma_2 E_?')= \dim\ker(E_?'\gamma_2)\\
               &= \dim\ker(\gamma_1 E_?'\gamma_2)= \dim\ker(\gamma_1\gamma_2),
\end{split}\end{equation}
 where we used that $\dim\ker A=\dim\ker A^\dag$ for any operator $A$ and that 
 $\gamma_1(\id- E_?')\gamma_2= \gamma_1(\id- E_?)\gamma_2 =0$.
\end{proof}

%------------------------------------------------------------------------------%
\subsection{Orthogonal subspaces}
An important consequence of the first part of Theorem~\ref{t4134} is the 
 following

\begin{lemma}\label{l8826}
Let $\mathcal E= (E_1, E_2, E_?)$ be an optimal USD measurement for a pair of 
 weighted density operators $\mathcal S= (\gamma_1, \gamma_2)$.
Suppose that $\Pi$ is a projector with $\Pi\Hilbert \subset
 (\ker\gamma_1\cap \supp \mathcal S)$.

Then $E_1\Pi= 0$ if and only if $E_2\Pi = \Pi$.
\end{lemma}
\begin{proof}
The ``if'' part follows directly from $0\le \Pi E_?\Pi= - \Pi E_1 \Pi$.
For the converse we have $\supp E_? \supset E_?\Pi\Hilbert =(\Pi- 
 E_2\Pi)\Hilbert \subset \ker\gamma_1$ and thus due to Theorem~\ref{t4134}, 
 $E_?\Pi\Hilbert\subset \ker S$.
But since $\ker S$ is orthogonal to $\Pi\Hilbert$, we have $\Pi E_?  \Pi= 0$.
Thus $0= E_?\Pi= \Pi- E_2\Pi$.
\end{proof}

In particular let $\Sigma_2$ denote the projector onto $\ker \gamma_1 \cap 
 \supp \gamma_2$.
Then necessarily for any USD measurement $E_1\Sigma_2= 0$ and hence by virtue 
 of Lemma~\ref{l8826}, for any optimal measurement $E_2\Sigma_2= \Sigma_2$ 
 holds.
With $\Sigma_1$ denoting the projector onto $\ker \gamma_2\cap \supp \gamma_1$ 
 we obtain $E_1\Sigma_1= \Sigma_1$ in an analogous way.
These observations are at the core of the following
\begin{proposition}[cf.\ Theorem~2 in Ref.~\cite{Raynal:2003PRA}]\label{p2770}
Let $\mathcal S= (\gamma_1, \gamma_2)$ be a pair of weighted density operators.
Denote by $\Pi_\mathrm{skew}$ the projector onto $(\ker\gamma_1+ 
 \supp\gamma_2)\cap (\ker\gamma_2+ \supp\gamma_1)$ and write $\mathcal 
 S^\mathrm{skew}=(\Pi_\mathrm{skew}\gamma_1\Pi_\mathrm{skew}, 
 \Pi_\mathrm{skew}\gamma_2\Pi_\mathrm{skew})$ for the projected pair.
Let $E_?$ and $E_?^\mathrm{skew}$ be two operators satisfying 
 $E_?^\mathrm{skew}= E_?+ (\id- \Pi_\mathrm{skew})$.
 
Then $E_?$ is an optimal and proper USD measurement for $\mathcal S$, if and 
 only if $E_?^\mathrm{skew}$ is an optimal and proper USD measurement for 
 $\mathcal S^\mathrm{skew}$.

In this case, the \emph{failure probability} for $\mathcal E$ of $\mathcal S$ 
 is the same as for $\mathcal E^\mathrm{skew}$ of $\mathcal S^\mathrm{skew}$,
\begin{equation}
 \tr(\gamma_1+\gamma_2)-P_\mathrm{succ}(\mathcal E,\mathcal S)=
 \tr[\Pi_\mathrm{skew}(\gamma_1+\gamma_2)]- P_\mathrm{succ}(\mathcal 
 E^\mathrm{skew},\mathcal S^\mathrm{skew}).
\end{equation}
\end{proposition}
\noindent (Note, that $(\ker\gamma_1+ \supp\gamma_2)\cap (\ker\gamma_2+ 
 \supp\gamma_1)$ is the orthocomplement of $(\Sigma_1+ \Sigma_2)\Hilbert$.
For the projected pair $\mathcal S^\mathrm{skew}$, the spaces 
 $\supp(\Pi_\mathrm{skew}\gamma_1\Pi_\mathrm{skew})$ and 
 $\supp(\Pi_\mathrm{skew}\gamma_2\Pi_\mathrm{skew})$ are skew, where two spaces 
 $\mathscr A$ and $\mathscr B$ are called \emph{skew}, if $\mathscr A \cap 
 \mathscr B^\perp= \{0\}= \mathscr B \cap \mathscr A^\perp$.)
\begin{proof}
Due to the discussion leading to the Proposition, for any optimal measurement 
 $E_?$ we have $E_?\Pi_\mathrm{skew}= E_?$ and hence $\Pi_\mathrm{skew}(\id- 
 E_?)\Pi_\mathrm{skew}=\id- E_?^\mathrm{skew}$.
It follows that $\id- E_?^\mathrm{skew}\ge 0$ and that 
 $\Pi_\mathrm{skew}\gamma_1\Pi_\mathrm{skew}(\id- 
 E_?^\mathrm{skew})\Pi_\mathrm{skew}\gamma_2\Pi_\mathrm{skew}= 0$.
Furthermore we find due to $\Pi_\mathrm{skew}\Hilbert\supset \ker \mathcal S$ 
 that
\begin{equation}\label{e2651}
 \ker \mathcal S^\mathrm{skew}=
 \ker[(\gamma_1+\gamma_2)\Pi_\mathrm{skew}]=
 \ker\Pi_\mathrm{skew} \oplus \ker \mathcal S,
\end{equation}
 where both terms in the direct sum are orthogonal.
This shows that $E_?^\mathrm{skew}$ acts as a projector onto $\ker\mathcal 
 S^\mathrm{skew}$.
Since obviously $E_?^\mathrm{skew}\ge 0$ we have shown that $E_?^\mathrm{skew}$ 
 is a proper USD measurement for $\mathcal S^\mathrm{skew}$.
The converse, namely that $E_?= E_?^\mathrm{skew}- (\id -\Pi_\mathrm{skew})$ is 
 a proper USD measurement of $\mathcal S$, in fact holds for any proper 
 measurement $E_?^\mathrm{skew}$ of $\mathcal S^\mathrm{skew}$.
This follows from Eq.~\eqref{e2651} and by noticing that for $\mathcal 
 E^\mathrm{skew}=(E_1^\mathrm{skew}, E_2^\mathrm{skew}, E_?^\mathrm{skew})$ the 
 measurement defined by $E_?$ is given by $\mathcal E=(E_1^\mathrm{skew}+ 
 \Sigma_1, E_2^\mathrm{skew}+ \Sigma_2, E_?)$.

In order to show that given $E_?$, the measurement $E_?^\mathrm{skew}$ is 
 optimal, suppose, that $E_?^{\mathrm{skew}\prime}$ is proper and has a higher 
 success probability than $E_?^\mathrm{skew}$.
Then it is easy to see that $E_?'= E_?^{\mathrm{skew}\prime}- (\id 
 -\Pi_\mathrm{skew})$ would yield a higher success probability for $\mathcal S$ 
 than $E_?$, in contradiction to the assumption.

On the other hand, since $E_?\Pi_\mathrm{skew}= E_?$, any optimal and proper 
 $E_?$ minimizes $\tr(E_? \Pi_\mathrm{skew}(\gamma_1+ 
 \gamma_2)\Pi_\mathrm{skew})$.
But this is minimal for optimal $E_?^\mathrm{skew}$, since $E_?= 
 E_?^\mathrm{skew} \Pi_\mathrm{skew}$.
\end{proof}

Proposition~\ref{p18391} and Proposition~\ref{p2770} can be used independently 
 from each other, in contrast to the original result in 
 Ref.~\cite{Raynal:2003PRA}.
Proposition~\ref{p18391} and Proposition~\ref{p2770} provide a method to obtain 
all optimal measurements\footnote{
In the original work \cite{Raynal:2003PRA} it was only shown that one 
 \emph{may} choose the measurements in that specific way.
Here we showed that all optimal measurements \emph{must} have this structure.
} for a given pair $\mathcal S$ by considering a different pair $\mathcal S'$ 
 where $\dim \supp\mathcal S \ge \dim \supp\mathcal S'$.
This is in particular useful, if $\dim \supp\mathcal S' \le 4$, since in 
 Sec.~\ref{s17133} we will provide an analytical solution for any such pair.
If $\dim \supp\mathcal S' \le 2$, then the general solution can already be 
  obtained due to the result by Jaeger and Shimony \cite{Jaeger:1995PLA}.
Also the pair $\mathcal S'$ might possess a two-dimensional common block 
  diagonal structure which was not present in the original pair $\mathcal S$ 
  and allows a solution of the problem (cf.\ Ref.~\cite{Bergou:2006PRA}; for a 
  simple criterion in order to detect such structures, cf.\ 
  Ref.~\cite{Kleinmann:2007JPA}).
Apart from that, using both propositions all optimal measurements can be found 
 by just considering pairs of states which do not possess any orthogonal (like 
 $\supp\gamma_1\cap \ker\gamma_2$) or parallel ($\supp\gamma_1\cap 
 \supp\gamma_2$) components.

The following property simplifies actual calculations.
\begin{lemma}\label{l2980}
With the notations of Proposition~\ref{p18391} and Proposition~\ref{p2770} let 
 $\tau_\nparallel$ denote the (non-linear) mapping from $\mathcal S$ to 
 $\mathcal S^\nparallel$ and analogously $\tau_\mathrm{skew}$ the mapping from 
 $\mathcal S$ to $\mathcal S^\mathrm{skew}$.

Then $\tau_\nparallel\circ \tau_\nparallel= \tau_\nparallel$,
 $\tau_\mathrm{skew}\circ \tau_\mathrm{skew}= \tau_\mathrm{skew}$ and
 $\tau_\mathrm{skew}\circ \tau_\nparallel= \tau_\nparallel\circ 
  \tau_\mathrm{skew}$.
\end{lemma}
\begin{proof}
We abbreviate $\tau[\mathcal S]_\mu$ for $\gamma_\mu$ after the application of 
 $\tau$, i.e., $(\tau[\mathcal S]_1, \tau[\mathcal S]_2)= \tau[\mathcal S]$.
One verifies
\begin{equation}
 \supp \tau_\nparallel[\mathcal S]_\mu=
  \supp\gamma_\mu\cap \supp \Pi_\nparallel,
\end{equation}
and
\begin{equation}\begin{split}
 \supp \tau_\mathrm{skew}[\mathcal S]_\mu&=
  \supp (\id- \Sigma_\mu)\gamma_\mu (\id- \Sigma_\mu)\\
  &=\supp \gamma_\mu \cap \ker\Sigma_\mu.
\end{split}\end{equation}

From the first equation we immediately get $\ker \tau_\nparallel[\mathcal S]_1
 + \ker \tau_\nparallel[\mathcal S]_2= \Hilbert$, i.e., $\tau_\nparallel$ is 
 acts as identity on $\tau_\nparallel[\mathcal S]$.
In order to show that $\tau_\mathrm{skew}$ is idempotent one verifies that 
 $\supp \tau_\mathrm{skew}[\mathcal S]_1 \cap \ker \tau_\mathrm{skew}[\mathcal 
 S]_2= \{0\}$.

Due to $\supp \tau_\nparallel[\mathcal S]_1 \cap \ker \tau_\nparallel[\mathcal 
 S]_2= \supp\gamma_1 \cap \ker\gamma_2$ it follows that
\begin{equation}
 (\tau_\mathrm{skew}\circ \tau_\nparallel) [\mathcal S]=
  (\Xi \gamma_1 \Xi, \Xi \gamma_2 \Xi),
\end{equation}
 where $\Xi= \Pi_\mathrm{skew}\Pi_\nparallel \equiv 
 \Pi_\nparallel\Pi_\mathrm{skew}$.
Analogously due to $\ker \tau_\mathrm{skew}[\mathcal S]_1+ \ker 
 \tau_\mathrm{skew}[\mathcal S]_2= \ker\gamma_1+ \ker\gamma_2$ we have
\begin{equation}
 (\tau_\nparallel\circ \tau_\mathrm{skew}) [\mathcal S]=
  (\Xi \gamma_1 \Xi, \Xi \gamma_2 \Xi),
\end{equation}
 and thus the third assertion holds.
\end{proof}
As an important consequence one can apply the mappings $\tau$ in any order and 
 in particular due to $(\tau_\mathrm{skew} \circ \tau_\nparallel)^{\circ 2}= 
 \tau_\mathrm{skew}\circ\tau_\nparallel$, a second application of both mappings 
 is never necessary.

The action of $\tau_\nparallel$ on $\mathcal S=(\gamma_1, \gamma_2)$ is 
 non-trivial, if and only if $\rank(\gamma_1+ \gamma_2)< \rank\gamma_1+ 
 \rank\gamma_2$.
Similarly, the action of $\tau_\mathrm{skew}$ is non-trivial if and only if 
 $\rank\gamma_1> \rank\gamma_1\gamma_2$ or $\rank\gamma_2>
 \rank\gamma_1\gamma_2$.
We call a pair of states $\mathcal S$ \emph{strictly skew}, if 
 $(\tau_\mathrm{skew}\circ\tau_\nparallel)[\mathcal S]= \mathcal S$.

Let us briefly mention a convenient way to construct the mapping 
 $\tau_\mathrm{skew}\circ \tau_\nparallel$.
As shown in the proof of Lemma~\ref{l2980}, we can write
\begin{equation}
 \mathcal S'\equiv (\tau_\mathrm{skew}\circ \tau_\nparallel)[\mathcal S]=
 (\Xi\gamma_1\Xi, \Xi\gamma_2\Xi),
\end{equation}
 with $\Xi=\id- \Pi_\parallel- \Sigma_1- \Sigma_2$.
Now let $(\ket{s_{1i}})$ and $(\ket{s_{2j}})$ be Jordan bases  (cf.\ 
 Appendix~\ref{a32712}) of $\supp\gamma_1$ and $\supp\gamma_2$.
Then $\Pi_\parallel=\sum_{i\in \mathcal X} \pr{s_{1i}}$ and $\Sigma_\mu= 
 \sum_{k\in \mathcal Y_\mu} \pr{s_{\mu k}}$, with $\mathcal X= \{ k\mid 
 \bracket{s_{1k}}{s_{2k}}= 1\}$, $\mathcal Y_1= \{ i\mid \forall j\colon 
 \bracket{s_{1i}}{s_{2j}}= 0 \}$, and $\mathcal Y_2= \{ j\mid \forall i\colon 
 \bracket{s_{1i}}{s_{2j}}= 0 \}$.

Summarizing Proposition~\ref{p18391} and Proposition~\ref{p2770}, if
 $\mathcal E'= (E_1', E_2', E_?')$ is an optimal and proper USD measurement of 
 $\mathcal S'=(\tau_\mathrm{skew}\circ \tau_\nparallel)[\mathcal S]$, then 
 $\mathcal E= (E_1'+ \Sigma_1, E_2'+ \Sigma_2, E_?'- \Sigma_1- \Sigma_2)$ is an 
 optimal and proper USD measurement of $\mathcal S$.
The optimal success probability computes to
\begin{equation}
 P_\mathrm{succ}(\mathcal E; \mathcal S)=
  P_\mathrm{succ}(\mathcal E'; \mathcal S')+ \tr[(\Sigma_1+ \Sigma_2) 
 (\gamma_1+ \gamma_2)].
\end{equation}
 
%------------------------------------------------------------------------------%
\subsection{Classification of USD measurements}\label{s12452}
We want to introduce a classification of the different types of optimal 
 measurements for USD.
Given the dimension of $\supp\mathcal S$, the classification is according to 
 the rank of the measurement operators.
For a Hilbert space of dimension $d$, we consider the optimal and proper USD 
 measurements $\mathcal E= (E_1, E_2, E_?)$ for pairs of weighted density 
 operators $\mathcal S= (\gamma_1, \gamma_2)$.
We restrict the analysis to the case, where $\tau_\mathrm{skew}$ and 
 $\tau_\nparallel$ act as identity on $\mathcal S$, i.e., to the case of 
 \emph{strictly skew} pairs.
Then $\rank \gamma_1\gamma_2= \rank \gamma_2= \rank\gamma_1\equiv r$ and 
 $\dim\ker\mathcal S= d-2r$ holds.
All optimal measurements with $\rank E_1= e_1$ and $\rank E_2= e_2$ will 
 be considered as one \emph{type} of measurement, denoted by $(e_1,e_2)$.
As we will see in subsequent sections, the construction method of the optimal 
 measurement mainly depends on the type of the measurement.
The symmetry of USD for exchanging the label of $\gamma_1$ and $\gamma_2$ makes 
 it only necessary to develop a construction procedure for the case where e.g.\ 
 $e_1\le e_2$.
Thus a measurement \emph{class} $[a,b]$ with $a\le b$ denotes both measurement 
 types $(a,b)$ and $(b,a)$.
We now count the number of measurement types and measurement classes.

Since we consider proper measurements, we have $\supp E_1\cap \supp E_2= \{0\}$ 
 and hence $e_1+ e_2= \rank (E_1+ E_2)$ and $e_\mu\le r$.
Let us denote by $\delta$ the dimension of the projective part of $E_?$, i.e, 
  $\delta= \dim\ker(\id- E_?)$.
Then $e_1+ e_2 + \delta= d$ and $\delta\le \rank E_?$.
From Theorem~\ref{t4134} we have that $\rank E_?= r+(d-2r)$.
On the other hand, at least $\ker(\id- E_?)\supset \ker\mathcal S$, i.e., 
 $\delta\ge d-2r$.
In summary we arrive at the constraints
\begin{equation}\label{e6632}
 e_1 \le r,\quad e_2\le r, \mx{and} r\le e_1+ e_2.
\end{equation}
From the situation where $\gamma_1$ and $\gamma_2$ have a two-dimensional 
 block diagonal structure, one can see that for any possible $e_1$ and $e_2$ 
 which satisfy the constraints in Eq.~\eqref{e6632}, one can find a pair 
 $\mathcal S=(\gamma_1,\gamma_2)$ such that an optimal measurement is of the 
 type $(e_1,e_2)$.

Counting the possible combinations to satisfy the conditions in 
 Eq.~\eqref{e6632}, one finds
\begin{equation}
 \#\mathrm{types}= \tfrac 12 (r+1)(r+2)
 \mx{and}
 \#\mathrm{classes}= \left\lfloor (\tfrac r2+1)^2 \right\rfloor,
\end{equation}
 where $\#\mathrm{types}$ denotes the number of measurement types and 
 $\#\mathrm{classes}$ the number of measurement classes.
Here we used the floor function, $\lfloor x \rfloor= \max\{ k\in \mathbb Z \mid 
 k\le x\}$.

Measurements of the type $(e_1,e_2)$ with $e_1+ e_2=r$ actually are von-Neumann 
 measurements.
(Obviously there are always $r+1$ such measurement types.)
This can be seen, since then $d-r= \rank E_?\ge \delta= d-e_1-e_2= d-r$, i.e., 
 $\rank E_?= \dim\ker(\id- E_?)$ and hence $E_?$ is projective.
But then $\tr E_1+\tr E_2= \tr(\id- E_?)= r= e_1+ e_2$ holds and hence all 
 eigenvalues of $E_1$ and $E_2$ are either $1$ or $0$.
This proves the assertion.

As we will see in Sec.~\ref{s194} and Sec.~\ref{s3330}, an analytic expression 
 for the optimal measurement is only known for the class $[r,r]$ and the 
 special von-Neumann class $[0,r]$.
These classes may occur for any $r\ge 1$ and thus in particular solve the 
 two-dimensional case ($r=1$) and ``half'' of the four-dimensional case 
 ($r=2$).
The remaining two classes (one of which is von-Neumann) in four dimensions are 
 solved in Sec.~\ref{Sec-DeltaOneDim} and Sec.~\ref{s25151}.

% vim:ft=tex ==================================================================%
\section{The optimality conditions by Eldar, Stojnic \& Hassibi}
Eldar, Stojnic, and Hassibi provided in Ref.~\cite{Eldar:2004PRA} necessary and 
 sufficient conditions for the optimality of a USD measurement\footnote{Indeed 
 Eldar \etal\ proved conditions for the optimality of a USD measurement for an 
 arbitrary number of states.}:
\begin{theorem}[Eldar, Stojnic \& Hassibi \cite{Eldar:2004PRA}]\label{t18481}
Let $\mathcal E=(E_1,E_2,E_?)$ be a proper USD measurement for a pair of 
 weighted density operators $\mathcal S=(\gamma_1, \gamma_2)$.
Denote by $\Lambda_1$ the projector onto $\ker\gamma_2\cap \supp\mathcal S$ and 
 by $\Lambda_2$ the projector onto $\ker\gamma_1\cap \supp\mathcal S$.
This measurement is optimal, if and only if one can find an operator $Z$ such 
 that for $\mu= 1,2$,
\begin{subequations}
\begin{gather}
  Z \ge 0,\quad ZE_?= 0,\\
  \Lambda_\mu(Z- \gamma_\mu) \Lambda_\mu\ge 0, \mx{and}
  \Lambda_\mu(Z- \gamma_\mu) E_\mu=0.\label{e16127b}
\end{gather}
\end{subequations}
\end{theorem}
In Ref.~\cite{Eldar:2004PRA}, this statement was only proved for the case 
 $\ker\mathcal S= \{0\}$.
However, the generalization presented in Theorem~\ref{t18481} follows 
 immediately from the original statement.

In Theorem~\ref{t18481} necessary and sufficient conditions for optimality 
 where presented.
However they are not operational, as the existence or non-existence of $Z$ is 
 difficult to prove.
We show in Appendix~\ref{s18521}, that the unknown operator $Z$ can be 
 eliminated, and the above conditions can be re-expressed as follows:
\begin{corollary}\label{c24797}
With the preliminaries and notations as in Theorem~\ref{t18481}, a proper 
 measurement $\mathcal E$ of $\mathcal S$ is optimal if and only if
\begin{subequations}
\label{e4736}
\begin{gather}
  \label{e4736a}
  (\Lambda_1- \Lambda_2) E_? (\gamma_2- \gamma_1) E_? (\Lambda_1+\Lambda_2)\ge 
  0 \\
  \label{e4736b}
  (\Lambda_1- \Lambda_2) E_? (\gamma_2- \gamma_1) E_?  (\id- E_?)=0.
\end{gather}
\end{subequations}
\end{corollary}
The conditions for an optimal USD measurement are now expressed as a series of 
 equations and positivity conditions on only $E_?$.
Remember the fact that $E_?$ already completely determines a USD measurement 
 (cf.\ Proposition~\ref{p6437}).

The first condition in the above Corollary~\ref{c24797}, Eq.~\eqref{e4736a}, 
 relies on the fact, that a positive semi-definite operator in particular has 
 to be self-adjoint.
Thus, the condition in Eq.~\eqref{e4736a} is only a compact notation for the 
 three conditions
\begin{subequations}\label{e26860}
\begin{align}
 \Lambda_1 E_? (\gamma_2- \gamma_1) E_? \Lambda_1 &\ge 0\label{e26860a},\\
 \Lambda_2 E_? (\gamma_1- \gamma_2) E_? \Lambda_2 &\ge 0\label{e26860b},\\
 \Lambda_1 E_? (\gamma_2- \gamma_1) E_? \Lambda_2 &= 0  \label{e26860c}.
\end{align}
\end{subequations}
(Obviously these conditions are sufficient for Eq.~\eqref{e4736a}.
The necessity follows from multiplication of Eq.~\eqref{e4736a} by $Q_\mu^\dag$ 
 from the left and $Q_\nu$ from the right.
Here $Q_\mu$ are the oblique projectors as defined in the proof of 
 Proposition~\ref{p6437}.)

The second equation, Eq.~\eqref{e4736b}, in Corollary~\ref{c24797} makes a 
 statement about the projective part of $E_?$.
This is the content of the following
\begin{lemma}\label{l17322}
Let $E_?$ be an optimal and proper USD measurement of a pair of weighted 
 density operators $\mathcal S=(\gamma_1, \gamma_2)$ with $\supp\gamma_1\cap 
 \supp\gamma_2= \{0\}$.
Denote by $\Pi_?$ the projector onto $\supp E_?$ and by $\Delta$ the projector 
 onto $\ker(\id- E_?)$.

Then $E_?(\gamma_2- \gamma_1)E_?= \Pi_?(\gamma_2- \gamma_1) \Pi_?= \Delta 
 (\gamma_2- \gamma_1)\Delta$.

\end{lemma}
\begin{proof}
Let $\Pi_\perp$ denote the projector onto $\ker\mathcal S$.
Then due to $\supp\gamma_1\cap \supp\gamma_2= \{0\}$, we have $(\Lambda_1- 
 \Lambda_2)\Hilbert= (\id- \Pi_\perp)\Hilbert$.
Due to $\Pi_\perp E_? \gamma_\mu= 0$, the optimality condition in 
 Eq.~\eqref{e4736b} hence reads $E_?(\gamma_2- \gamma_1)E_?(\id- E_?)= 0$ or
\begin{equation}
 E_?(\gamma_2- \gamma_1)E_?= E_?(\gamma_2- \gamma_1){E_?}^2.
\end{equation}
For the first equality we multiply this equation from the right by the inverse 
 (on its support) of $E_?$ and in a second step from the left and obtain the 
 equations
\begin{subequations}
\begin{align}
 E_?(\gamma_2-\gamma_1)\Pi_?&= E_?(\gamma_2-\gamma_1)E_?,\\
 \Pi_?(\gamma_2-\gamma_1)\Pi_?&= \Pi_?(\gamma_2-\gamma_1)E_?.
\end{align}
\end{subequations}
Since the right hand side of the first equation is self-adjoint, the assertion 
 follows.

For the second equality we have $E_?\Delta= \Delta$ and $N\equiv E_?(\id 
 -\Delta)= E_?- \Delta$ with $N\Delta= 0$.
Thus $E_?(\id- E_?)= N(\id -N)$.
But $\ker(\id- N)= \ker(\id- E_?+ \Delta)= \{0\}$ and hence the optimality 
 condition in Eq.~\eqref{e4736b} reads $E_?(\gamma_2-\gamma_1)N= 0$.
Thus
\begin{subequations}
\begin{align}
 E_?(\gamma_2- \gamma_1)E_?&= E_?(\gamma_2- \gamma_1)\Delta\\
 \Delta(\gamma_2- \gamma_1)E_?&= \Delta(\gamma_2- \gamma_1)\Delta
\end{align}
\end{subequations}
 holds, where in the second step we multiplied the first equation by $\Delta$ 
 from the left.
\end{proof}

Lemma~\ref{l17322} is the key to prove the uniqueness of the optimal and proper 
 USD measurement, since due to the identity in Eq.~\eqref{e5125} we have seen 
 that any USD measurement is solely defined by $E_?(\gamma_2- \gamma_1)E_?$.
Hence in the case of $\supp\gamma_1\cap \supp\gamma_2= \{0\}$, the optimal and 
 proper USD measurement can be uniquely determined, given $\Pi_?$, the 
 projector onto the support of $E_?$.
But since the set of optimal and proper USD measurements of $\mathcal S$ is by 
 virtue of Proposition~\ref{p18391} equal to the set of optimal and proper USD 
 measurements of $\mathcal S^\nparallel$, having $\supp\gamma_1\cap 
 \supp\gamma_2= \{0\}$ (cf.\ also Lemma~\ref{l2980}), it remains to show that 
 the support of $E_?$ is unique.
This follows from the fact that the rank of $E_?$ is fixed by virtue of 
 Theorem~\ref{t4134}, together with the convexity of optimal and proper 
 measurements.
Namely, for any two optimal and proper USD measurements $E_?$ and $\tilde E_?$, 
 also $\frac12(E_?+ \tilde E_?)$ is an optimal and proper USD measurement.
But since $E_?$ and $\tilde E_?$ are positive semi-definite, $\rank (E_?+ 
 \tilde E_?)= \rank E_?= \rank \tilde E_?$ can only hold if $\supp E_?= \supp 
 \tilde E_?$.
Thus we have proved the following
\begin{proposition}\label{p14866}
For a given pair of weighted density operators, there exists exactly one 
 optimal and proper USD measurement.
\end{proposition}

% vim:ft=tex ==================================================================%
\section{Two special classes of optimal measurements}
%------------------------------------------------------------------------------%
\subsection{Single state detection}\label{s194}
For certain pairs of weighted density operators $\mathcal S=(\gamma_1, 
 \gamma_2)$ it may be advantageous to choose a measurement with 
 $\tr(E_1\gamma_1)= 0$, e.g.\ if $\tr(\gamma_1)$ is much smaller than 
 $\tr(\gamma_2)$.
We refer to this situation as \emph{single state detection} of $\gamma_2$.
In the classification scheme proposed in Sec.~\ref{s12452}, the single state 
 detection measurements can be identified with the class $[0,r]$ (where $r= 
 \rank \gamma_1\gamma_2$).

For a proper measurement, $\tr(E_1\gamma_1)= 0$ can only hold if already $E_1= 
 0$.
If the measurement is optimal then due to Lemma~\ref{l8826}, $E_1= 0$ implies 
 $E_2= \Lambda_2$.
(The projectors $\Lambda_\mu$ were defined in Theorem~\ref{t18481}.)
It follows that $E_?=\id- E_1- E_2= \id- \Lambda_2$ is a projector and hence 
 satisfies the optimality condition in Eq.~\eqref{e4736b}.
Thus the measurement is optimal if and only if Eq.~\eqref{e4736a} holds, i.e.,
\begin{equation}
 \Lambda_1E_?(\gamma_2- \gamma_1)E_?\Lambda_1\ge 0.
\end{equation}
Let us now assume that $\supp\gamma_1\cap \supp\gamma_2= \{0\}$.
Then $(\id- \Lambda_2)\Lambda_1\Hilbert= \gamma_1\Hilbert$ and we arrive at the 
 following
\begin{proposition}\label{p6402}
Let $\mathcal E=(E_1,E_2,E_?)$ be an optimal USD measurement for a pair of 
 weighted density operators $\mathcal S=(\gamma_1, \gamma_2)$ with 
 $\supp\gamma_1\cap \supp\gamma_2= \{0\}$.
Then $\tr(E_1\gamma_1)= 0$ if and only if $\gamma_1(\gamma_2- 
 \gamma_1)\gamma_1\ge 0$.

In this case the success probability is given by $P_\mathrm{succ}(\mathcal 
 E;\mathcal S)= \tr(\Lambda_2\gamma_2)$, and if $\mathcal E$ is proper, then 
 $\mathcal E=(0,\Lambda_2,\id- \Lambda_2)$.
($\Lambda_2$ is the projector onto $\ker\gamma_1\cap \supp\mathcal S$.)
\end{proposition}
\begin{proof}
Assume, that $\mathcal E$ is optimal and satisfies $\tr(E_1\gamma_1)= 0$.
Then also for the corresponding proper measurement $\mathcal 
 E'=(E_1',E_2',E_?')$ (cf.\ Sec.~\ref{s20689}) we have $\tr(E'_1\gamma_1)= 0$ 
 and hence $\gamma_1(\gamma_2- \gamma_1)\gamma_1\ge 0$ follows.
For the contrary, we have already shown that the if $\gamma_1(\gamma_2- 
 \gamma_1)\gamma_1\ge 0$ holds, then the proper measurement $\mathcal 
 E=(0,\Lambda_2,\id- \Lambda_2)$ is an optimal measurement.
But due to Proposition~\ref{p14866}, this is the only optimal and proper 
 measurement.
Let now $\bar{\mathcal E}=(\bar E_1,\bar E_2,\bar E_?)$ be some optimal 
 measurement, that is not proper.
Then if the projection $E_1$ of $\bar E_1$ onto $\supp\mathcal S$ satisfies 
 $\tr(E_1\gamma_1)= 0$, then necessarily also $\tr(\bar E_1 \gamma_1)= 0$ 
 holds.
\end{proof}

Let us consider the situation, where the success probability for the states 
 $\rho_1$ and $\rho_2$ (both having unit trace) is analyzed in dependence of 
 the \apriori\ probability $0< p_1< 1$ of the state $\rho_1$, while the 
 \apriori\ probability of $\rho_2$ is $p_2=1- p_1$.
Then the optimality condition in Proposition~\ref{p6402} is satisfied, if and 
 only if for any $\ket\varphi \in \supp\rho_1$,
\begin{equation}
 (1-p_1)\bra\varphi\rho_2\ket\varphi
     \ge p_1\bra\varphi\rho_1\ket\varphi.
\end{equation}
If there exists a $\ket\varphi\in \supp\rho_1\cap \ker\rho_2$ with 
 $\ket\varphi\ne 0$, then this condition cannot be satisfied for any $p_1>0$.
But if we assume $\supp\rho_1\cap \ker\rho_2=\{0\}$, single state detection of 
 $p_2\rho_2$ is optimal if and only if $0< p_1 \le \ell_1$, where $\ell_1$ is 
 given by (with $\ket\varphi\in \supp\rho_1$ and $\bracket\varphi\varphi=1$)
\begin{equation}
 \ell_1= \min_{\ket\varphi}\left\{
         \frac{\bra\varphi         \rho_2 \ket\varphi}{
               \bra\varphi(\rho_1+ \rho_2)\ket\varphi} \right\}=
          \frac{\lambda_1}{1+ \lambda_1},
\end{equation}
 where ($\sqrt{\rho_1}\,^-$ denotes the inverse of $\sqrt{\rho_1}$ on its 
 support)
\begin{equation}
 \lambda_1=\min_{\ket\varphi} \bra\varphi\sqrt{\rho_1}\,^-\rho_2
                                   \sqrt{\rho_1}\,^-\ket\varphi.
\end{equation}
The minimum in the expression for $\lambda_1$ is given by the smallest 
 non-vanishing eigenvalue of the operator $\sqrt{\rho_1}\,^-\rho_2 
 \sqrt{\rho_1}\,^-$ (remember, that we assumed $\supp\rho_1\cap \ker\rho_2= 
 \{0\}$).
Note that $\lambda_1> 0$ and hence there always exists a finite parameter range 
 for $p_1$, where single state detection of $\gamma_2$ is optimal.

An analogous construction yields $\ell_2$, such that single state detection of 
 $\gamma_1$ is optimal if and only if $0< p_2 \le \ell_2$.

%------------------------------------------------------------------------------%
\subsection{Fidelity form measurement}\label{s3330}
An upper bound on the optimal success probability of USD was constructed by 
 Rudolph, Spekkens and Turner in Ref.~\cite{Rudolph:2003PRA}.
Let $\pr{\gamma_\mu}$ be a {\em purification} 
 \cite{Hughston:1993PLA,Bassi:2003PLA} of $\gamma_\mu$, i.e., a positive 
 semi-definite operator of $\rank 1$ acting on an extended Hilbert space 
 $\Hilbert\otimes \Hilbert_\mathrm{aux}$, such that the partial trace over 
 $\Hilbert_\mathrm{aux}$ yields back the original weighted density operator, 
 $\tr_\mathrm{aux} \pr{\gamma_\mu}= \gamma_\mu$.
Since the partial trace can be implemented by physical means, the optimal 
 unambiguous discrimination of $\mathcal S=(\gamma_1,\gamma_2)$ cannot have a 
 higher success probability than $\mathcal S^\mathrm{pur}= 
 (\pr{\gamma_1},\pr{\gamma_2})$.
But $\mathcal S^\mathrm{pur}$ is a pair of pure states, for which the optimal 
 success probability is known due to the result by Jaeger and Shimony 
 \cite{Jaeger:1995PLA}.
The map from $\mathcal S$ to $\mathcal S^\mathrm{pur}$, on the other hand, can 
 only be performed physically in very special situations 
 \cite{Kleinmann:2006PRA,Kleinmann:2007APB} and hence the success probability 
 of $\mathcal S^\mathrm{pur}$ in general only yields an upper bound.
This bound is strongly related to the Uhlmann fidelity 
 $\tr\abs{\sqrt{\rho_1}\sqrt{\rho_2}}$ of $\rho_1\equiv \gamma_1/\tr(\gamma_1)$ 
 and $\rho_2\equiv \gamma_2/\tr(\gamma_2)$ 
 \cite{Uhlmann:1976RMP,Jozsa:1994JMO}.
The Uhlmann fidelity is the largest overlap between any purification of both 
 states $\rho_1$ and $\rho_2$.
Due to this relation the bound was named \emph{fidelity bound} 
 \cite{Herzog:2005PRA}.
In Ref.~\cite{Herzog:2005PRA,Raynal:2005PRA}, necessary and sufficient 
 conditions for the fidelity bound to be optimal where shown and the optimal 
 measurement was constructed.
In this section we summarize and extend these results.
 
We continue to assume $\supp\gamma_1\cap \supp\gamma_2= \{0\}$.
Herzog and Bergou showed in Ref.~\cite{Herzog:2005PRA} that the fidelity bound 
 can be reached only if $E_?(\gamma_2- \gamma_1)E_?= 0$.
(From Corollary~\ref{c24797}, it is obvious that any such measurement is 
 optimal.)
Then due to Eq.~\eqref{e5125} we find that any measurement with $E_?(\gamma_2- 
 \gamma_1)E_?= 0$ is given by
\begin{equation}\label{e14310}
\begin{split}
 E_?&= \Pi_\perp+ (\gamma_1+ \gamma_2)^-\big\{\gamma_1\gamma_2+\gamma_2\gamma_1
      +\sqrt{\gamma_1}F_1\sqrt{\gamma_1}
      +\sqrt{\gamma_2}F_2\sqrt{\gamma_2}\big\}(\gamma_1+\gamma_2)^-\\
     &= \id- (\gamma_1+ \gamma_2)^-\big\{
                  \sqrt{\gamma_1}(\gamma_1- F_1)\sqrt{\gamma_1}
             + \sqrt{\gamma_2}(\gamma_2- F_2)\sqrt{\gamma_2}
             \big\}(\gamma_1+ \gamma_2)^-,
\end{split}
\end{equation}
 where we abbreviated $F_1= \sqrt{\sqrt{\gamma_1} \gamma_2 \sqrt{\gamma_1}}$ 
 and $F_2= \sqrt{\sqrt{\gamma_2} \gamma_1 \sqrt{\gamma_2}}$.
The converse is also true:
\begin{lemma}\label{l22262}
Let $\mathcal S= (\gamma_1, \gamma_2)$ be a pair of weighted density operators 
 with $\supp\gamma_1\cap \supp\gamma_2= \{0\}$ and let $\mathcal 
 E=(E_1,E_2,E_?)$ be a proper USD measurement of $\mathcal S$.
Then $E_?\gamma_2 E_?= E_? \gamma_1 E_?$ if and only if $E_?$ is given by 
 Eq.~\eqref{e14310}.
\end{lemma}
\begin{proof}
It remains to show the ``if'' part.
First we multiply the identity
\begin{equation}
 (\gamma_1+\gamma_2)(\gamma_1+\gamma_2)^-\gamma_1= \gamma_1
\end{equation}
 from left by $Q_1$ as defined in the proof of Proposition~\ref{p6437}, i.e., 
 $Q_1$ is the oblique projector from $\ker\gamma_2\cap \supp\mathcal S$ to 
 $\supp\gamma_1$.
We then obtain due to Lemma~\ref{l10547} (cf.\ Appendix~\ref{s26828}) that 
 $\gamma_1(\gamma_1+ \gamma_2)^-\gamma_1= \gamma_1$ and thus 
 $\gamma_2(\gamma_1+ \gamma_2)^-\gamma_1= 0$.
(One can show that $\gamma_1(\gamma_1+\gamma_2)^-= Q_1$.)
From the polar decomposition of $\sqrt{\gamma_1}\sqrt{\gamma_2}$, it 
 furthermore follows, that there exists a unitary transformation $U$, such that 
 $\sqrt{\gamma_1}\sqrt{\gamma_2} U= F_1$ (and hence $U 
 \sqrt{\gamma_1}\sqrt{\gamma_2}= F_2$, cf.\ also Ref.~\cite{Raynal:2005PRA}).
Thus
\begin{equation}
\begin{split}
 E_?\sqrt{\gamma_1}&=(\gamma_1+ \gamma_2)^-\{0+ \gamma_2 \sqrt{\gamma_1}+
                      \sqrt{\gamma_1}(\sqrt{\gamma_1}\sqrt{\gamma_2}U)+ 0\}\\
                   &=(\gamma_1+ \gamma_2)^-\{\sqrt{\gamma_2}(U\sqrt{\gamma_1}
                       \sqrt{\gamma_2})^\dag+ \gamma_1\sqrt{\gamma_2}\}U \\
                   &=E_?\sqrt{\gamma_2}U,
\end{split}
\end{equation}
i.e., we have $E_?\gamma_1E_?= E_?\gamma_2E_?$.
\end{proof}

We refer to the measurement characterized by Lemma~\ref{l22262} as 
 \emph{fidelity form measurement} due to the appearance of the operators $F_1$ 
 and $F_2$, which satisfy $\tr\abs{\sqrt{\gamma_1}\sqrt{\gamma_2}}=\tr F_1= \tr 
 F_2$.
According to the classification in Sec.~\ref{s12452}, the fidelity form 
 measurements are a (strict) superset of the measurement class $[r,r]$, where 
 $r=\rank \gamma_1 \gamma_2$.
(This can be seen from Lemma~\ref{l17322}: in the class $[r,r]$ we have 
 $\dim\ker(\id- E_?)= 0$ and hence in particular $E_?(\gamma_2-\gamma_1)E_?= 
 0$.)

Unfortunately, it is very rare that the operator given by Eq.~\eqref{e14310} is 
 part of a valid USD measurement.
The following Proposition states necessary and sufficient criteria.
\begin{proposition}[cf.\ Theorem~4 in Ref.~\cite{Raynal:2005PRA}]\label{p22024}
Let $\mathcal S=(\gamma_1, \gamma_2)$ be a pair of weighted density operators 
 with $\supp\gamma_1\cap \supp\gamma_2= \{0\}$.
Then there exists a proper USD measurement $\mathcal E=(E_1,E_2,E_?)$ of 
 $\mathcal S$ with $E_?$ given by Eq.~\eqref{e14310} if and only if $\gamma_1- 
 \sqrt{\sqrt{\gamma_1} \gamma_2 \sqrt{\gamma_1}}\ge 0$ and $\gamma_2- 
 \sqrt{\sqrt{\gamma_2} \gamma_1 \sqrt{\gamma_2}}\ge 0$.

If the measurement exists, it is optimal and the success probability is given 
 by $P_\mathrm{succ}(\mathcal E,\mathcal S)=\tr(\gamma_1+\gamma_2)- 2\tr \abs{ 
 \sqrt{\gamma_1}\sqrt{\gamma_2} }$.
\end{proposition}
(Note that $\tr(\gamma_1+\gamma_2)- 2\tr \abs{ \sqrt{\gamma_1}\sqrt{\gamma_2} 
 }$ is the square of the Bures distance 
 \cite{Bures:1969TAMS,Uhlmann:1976RMP,Jozsa:1994JMO} between $\gamma_1$ and 
 $\gamma_2$.)
\begin{proof}
Due to the properties shown in the proof of Lemma~\ref{l22262}, for $E_?$ given 
 by Eq.~\eqref{e14310} we have $\gamma_1(\id- E_?)\gamma_2= 0$ and 
 $\sqrt{\gamma_\mu}(\id- E_?)\sqrt{\gamma_\mu}= \gamma_\mu- F_\mu$.
Furthermore, one can write $E_?=\Pi_\perp+ AA^\dag$, with (cf.\ 
 Ref.~\cite{Raynal:2005PRA})
\begin{equation}
 A= (\gamma_1+ \gamma_2)^-(\sqrt{\gamma_1}+ \sqrt{\gamma_2} U) \sqrt{F_1},
\end{equation}
 where $U$ is a unitary transformation originating from the polar decomposition 
 $\sqrt{\gamma_1}\sqrt{\gamma_2} U= F_1$ (cf.\ proof of Lemma~\ref{l22262}).

Due to these properties, the necessary and sufficient conditions on $E_?$ shown 
 in Proposition~\ref{p6437} reduce to the assertion of the current Proposition.
\end{proof}

If the criterion in Proposition~\ref{p22024} is not satisfied, then the optimal 
 measurement cannot be of the form as given by Eq.~\eqref{e14310}.
Thus by virtue of Lemma~\ref{l22262}, $E_?(\gamma_2- \gamma_1)E_?\ne 0$ holds 
 and using Lemma~\ref{l17322} we have that $\ker(\id- E_?)\supsetneq 
 \ker\mathcal S$.
But due to Eq.~\eqref{e25118} it follows that $\ker(\id- E_?)$ contains at 
 least one vector either in $\supp\gamma_1$ or in $\supp\gamma_2$ (cf.\ 
 Corollary~1 in Ref.~\cite{Raynal:2007PRA}).

Similar to the discussion of the single state detection measurement in 
 Sec.~\ref{s194}, we ask for the values of the \apriori\ probability 
 $0<p_1<1$ of $\rho_1$, for which the fidelity form measurement is optimal.
The first condition in Proposition~\ref{p22024}, $\gamma_1- F_1 \ge 0$, is 
 satisfied if and only if for any $\ket\varphi\in \supp\gamma_1$,
\begin{equation}
  p_1\bra\varphi\rho_1\ket\varphi\ge \sqrt{p_1(1-p_1)} \bra\varphi R_1 
\ket\varphi
\end{equation}
 holds, where we abbreviated $R_1= \sqrt{\sqrt{\rho_1}\rho_2\sqrt{\rho_1}}$.
Thus $\gamma_1- F_1 \ge 0$ if and only if $m_1\le p_1 <1$, where $m_1$ is given 
 by
\begin{equation}
 m_1=\frac{\mu_1^2}{1+\mu_1^2},
\end{equation}
 with $\mu_1$ the maximal eigenvalue of $\sqrt{\rho_1}\,^- R_1 
 \sqrt{\rho_1}\,^-$.
With an analogous construction we get that $\gamma_2-F_2\ge 0$ if and only if 
 $m_2\le p_2\equiv 1-p_1 < 1$.
Then
\begin{equation}
 m_1\le p_1 \le 1-m_2
\end{equation}
 is the region where the fidelity form measurement is optimal.
Note, that this region is empty when $m_1+m_2>1$.

In summary, single state detection is optimal if and only if \{($\gamma_1^2- 
 F_1^2\le 0$) \emph{or} ($\gamma_2^2- F_2^2\le 0$)\}, while the fidelity form 
 measurement is optimal if and only if \{($\gamma_1-F_1\ge 0$) \emph{and} 
 ($\gamma_2-F_2\ge 0$\}.
The situations, where the optimal measurement is neither a single state 
 detection measurement nor a fidelity form measurement seems to be related to 
 the gap between ``$A\ge B$'' and ``$A^2\le B^2$'' for positive operators $A$ 
 and $B$.
In the pure state case, however, $A$ and $B$ are of rank $1$ and hence this gap 
 does not exist.
Indeed, in the pure state case, $\mu_1^2= \mu_2^2= \tr(\rho_1\rho_2)= 
 \lambda_1= \lambda_2$ (where $\lambda_\mu$ and $\ell_\mu$ were defined at the 
 end of Sec.~\ref{s194}).
Thus $\ell_1= m_1$ and $\ell_2= m_2$ and hence either the single state 
 detection or the fidelity form measurement is always optimal.
This is exactly the solution for the pure state case, as given by Jaeger and 
 Shimony in Ref.~\cite{Jaeger:1995PLA}.

% vim:ft=tex ==================================================================%
\section{Solution in four dimensions}\label{s17133}
In this section we reduce the candidates for an optimal and proper USD 
 measurement for the case where $\dim\supp\mathcal S= 4$ to a finite number.
These candidates are obtained by finding the real roots of a high-order 
 polynomial.

Due to Proposition~\ref{p18391} and Proposition~\ref{p2770}, it is sufficient 
 to discuss the case of strictly skew pairs $\mathcal S= (\gamma_1, \gamma_2)$ 
 with $\ker\mathcal S= \{0\}$, i.e.,
\begin{equation}\label{e6795}\begin{split}
 \supp\gamma_1\cap \supp\gamma_2=\{0\},&\quad
 \ker\gamma_1\cap \ker\gamma_2=\{0\},\\
 \supp\gamma_1\cap \ker\gamma_2=\{0\}, &\mx{and}
 \ker\gamma_1\cap \supp\gamma_2=\{0\}.
\end{split}\end{equation}
Then, following the discussion in Sec.~\ref{s12452}, there are six types of 
 optimal USD measurements which differ in the rank of the measurement operators 
 $E_1$ and $E_2$.
We list these possible types in Table~\ref{Tab-POV4dim}.
\begin{table}
\caption{\label{Tab-POV4dim}
Possible ranks of the measurement operators of the optimal USD measurement in 
 four dimensions (cf.\ also Sec.~\ref{s12452}).
}
\begin{center}\begin{tabular}{c|c|c|c|c|c}
$\rank E_1$ & $\rank E_2$ & $\rank E_?$ & $\dim\ker(\id- E_?)$ & class & cf.\ 
Sec.\\
\hline
0 & 2 & 2 & 2 & $[0,2]$ & \ref{s194}\\
1 & 2 & 2 & 1 & $[1,2]$ & \ref{Sec-DeltaOneDim}\\
2 & 2 & 2 & 0 & $[2,2]$ & \ref{s3330}\\
1 & 1 & 2 & 2 & $[1,1]$ & \ref{s25151}\\
2 & 1 & 2 & 1 & $[1,2]$ & \ref{Sec-DeltaOneDim}\\
2 & 0 & 2 & 2 & $[0,2]$ & \ref{s194}\\
\end{tabular}\end{center}
\end{table}

The measurements of the class $[0,2]$ and the class $[2,2]$ where already 
 extensively discussed in Sec.~\ref{s194} and Sec.~\ref{s3330}, respectively.
For the class $[1,2]$, the kernel of $(\id- E_?)$ is one-dimensional.
We will consider this class in Sec.~\ref{Sec-DeltaOneDim}.
In the remaining case, where $\rank E_1=1 = \rank E_2$, the measurement is a 
 von-Neumann measurement (cf.\ Sec.~\ref{s12452}).
An example of this kind of measurement was first found in 
 Ref.~\cite{Raynal:2007PRA}.
Sec.~\ref{s25151} is devoted for a general treatment of this class.

%------------------------------------------------------------------------------%
\subsection{The measurement class $[1,2]$}\label{Sec-DeltaOneDim}
In this class the dimension of $\ker(\id- E_?)$ is $1$.
According to Eq.~\eqref{e25118}, this kernel is either contained in 
 $\supp\gamma_1$ or in $\supp\gamma_2$.
Here we focus on $\ker(\id- E_?)\subset \supp\gamma_1$, i.e., to the 
 measurement type $(1,2)$; the case of $\ker(\id- E_?)\subset \supp\gamma_2$ 
 follows along same lines.
Thus there exists an orthonormal basis $(\ket\phi,\ket{\phi^\perp})$ of 
 $\supp\gamma_1$, such that
\begin{equation}\label{e18439}
 E_?=\nu \pr n +\pr\phi,
\end{equation}
 with $\ket n$ some normalized vector, orthogonal to $\ket\phi$ and $0<\nu <1$.

%..............................................................................%
\subsubsection{The necessary and sufficient conditions}\label{s9110}
Due to Proposition~\ref{p6437}, the operator in Eq.~\eqref{e18439} is a valid 
 USD measurement, if and only if $\gamma_1(\id -E_?)\gamma_2= 0$, i.e.,
\begin{equation}
 \gamma_1\pr{\phi^\perp}\gamma_2 = \nu \gamma_1 \pr n\gamma_2
\end{equation}
 holds.
This is equivalent to
\begin{equation}\label{Eq-2x1ab}
 \gamma_1\ket n= a\gamma_1\ket{\phi^\perp} \mx{and}
 \gamma_2\ket n= b\gamma_2\ket{\phi^\perp}.
\end{equation}
 where $a= \bracket{\phi^\perp}n$ and $b=(\nu \bracket n{\phi^\perp})^{-1}$.
Remember, that due to Theorem~\ref{t4134} we have $\ket n\notin \ker\gamma_1$, 
 i.e., $\bracket n {\phi^\perp}\neq 0$.
On the other hand, from $\nu<1$ it follows $a b^*>1$ and thus $\ket n\notin 
 \vspan\{\ket{\phi^\perp}\}$, i.e., $\ket n\notin \supp\gamma_1$.
Given $\ket{\phi^\perp}$, we choose the phase of $\ket n$ such that $\bracket n 
 {\phi^\perp}>0$.

The conditions in Corollary~\ref{c24797} can now be reduced to simple 
relations.
By multiplying Eq.~\eqref{e4736b} from the left with 
 $\pr{\phi^\perp}(\Lambda_1- \Lambda_2)^{-1}$, this first leads to
\begin{equation}
 \bra n \gamma_2-\gamma_1 \ket n= 0.
\end{equation}
Now it is straightforward to see that the conditions in Eq.~\eqref{e4736} are 
 equivalent to
\begin{align}\label{e488}
 E_?(\gamma_2- \gamma_1)\ket n&= 0, \quad\text{and}\\
 \label{Eq-Eldar1x2e3}
 \bra\phi \gamma_2-\gamma_1 \ket\phi&\ge 0.
\end{align}
By virtue of Eq.~\eqref{Eq-2x1ab}, we can re-express Eq.~\eqref{e488} in terms 
 of $\ket\phi$ and $\ket{\phi^\perp}$, yielding
\begin{align}\label{e25745}
 \sqrt{\bra{\phi^\perp} \gamma_1 \ket{\phi^\perp}}
  \bra{\phi^\perp} \gamma_2 \ket\phi&=
 \sqrt{\bra{\phi^\perp} \gamma_2 \ket{\phi^\perp}}
  \bra{\phi^\perp} \gamma_1 \ket\phi, \quad\text{and}\\\label{e21583}
 \frac ab&= \sqrt{
  \frac{\bra{\phi^\perp}\gamma_2\ket{\phi^\perp}}
       {\bra{\phi^\perp}\gamma_1\ket{\phi^\perp}}}.
\end{align}

The last equation enables us to construct $\nu \pr n$ from
\begin{equation}\label{e31410}\begin{split}
 \sqrt\nu\ket n&=\sqrt\nu(\gamma_1+\gamma_2)^{-1}(\gamma_1+\gamma_2) \ket n\\
     &= \sqrt\nu(\gamma_1+\gamma_2)^{-1}(a\,\gamma_1+ 
     b\,\gamma_2)\ket{\phi^\perp}\\
     &= K\ket{\phi^\perp},
\end{split}
\end{equation}
 where in the last step we used $\sqrt{ab\nu}= 1$ and we abbreviated
\begin{equation}
 K=(\gamma_1+\gamma_2)^{-1} \left(\sqrt{a/b}\,\gamma_1+
   \sqrt{b/a}\,\gamma_2\right).
\end{equation}
But due to Eq.~\eqref{e21583}, $\sqrt{a/b}$ is given in terms of $\ket\phi$ and 
 $\ket{\phi^\perp}$.
Note, that $K$ has full rank and hence ensures $\nu\equiv 
 \bra{\phi^\perp}K^\dag K \ket{\phi^\perp}>0$.

We summarize:\emph{
The optimal USD measurement is of type $(1,2)$ if and only if there exists an 
 orthonormal basis $(\ket\phi,\ket{\phi^\perp})$ of $\supp\gamma_1$, such that 
 the conditions in Eq.~\eqref{Eq-Eldar1x2e3}, in Eq.~\eqref{e25745}, and 
 $\bra{\phi^\perp}K^\dag K\ket{\phi^\perp}<1$ are satisfied.}

%..............................................................................%
\subsubsection{Construction of a finite number of candidates for $E_?$}
In the following we will show, that already Eq.~\eqref{e25745} reduces the 
 possible candidates of $\vspan\{\ket\phi\}$ to a finite number and hence the 
 remaining positivity conditions can be easily checked.
Eq.~\eqref{e25745} is a complex equation.
Thus the absolute value and the phase of the left hand side and the right hand 
 side have to be identical.
This leads to
\begin{gather}\label{Eq-OptDelta2Eq1}
 \bra\phi \gamma_2 \ket{\phi^\perp}
  \bra{\phi^\perp} \gamma_2 \ket\phi
  \bra{\phi^\perp} \gamma_1 \ket{\phi^\perp}=
 \bra\phi \gamma_1 \ket{\phi^\perp}
  \bra{\phi^\perp} \gamma_1 \ket\phi
  \bra{\phi^\perp} \gamma_2 \ket{\phi^\perp},
 \\\label{Eq-OptDelta2Eq2}
 \bra{\phi^\perp} \gamma_2 \ket\phi
 \bra\phi \gamma_1 \ket{\phi^\perp}\ge 0.
\end{gather}

Let $(\ket{s_1}, \ket{s_2})$ be an orthonormal basis of $\supp\gamma_1$, such 
 that $\gamma_1\ket{s_i}= g_{1i}\ket{s_i}$ with $g_{11}\ge g_{12}>0$.
We abbreviate $g_{2i}= \bra{s_i}\gamma_2\ket{s_i}> 0$, $g_\mu= g_{\mu 1}-g_{\mu 
 2}$, and $g_{23}= \bra{s_1} \gamma_2 \ket{s_2}.$
We ensure $g_{23}\ge 0$ by choosing a proper global phase of $\ket{s_2}$.
In the case $g_1= 0$ we use a basis where $g_{23}= 0$.

First we consider, whether $\ket{s_1}$ or $\ket{s_2}$ is a candidate for 
 $\ket\phi$.
In either case, Eq.~\eqref{e25745} can only be satisfied, if $g_{23}= 0$.
From Eq.~\eqref{Eq-Eldar1x2e3} we find that $\ket{s_1}$ is a candidate only if 
 $g_{21}\ge g_{11}$ and analogously, $\ket{s_2}$ only if $g_{22}\ge g_{12}$.

We now assume that neither of the above two cases is optimal.
Then any of the remaining bases (apart from global phases) are parametrized by
\begin{subequations}\label{e23160}
\begin{align}
 \ket\phi        &=(1+x^2)^{-\frac12}(\ket{s_1}+x\e{i\vartheta} \ket{s_2}),\\
 \ket{\phi^\perp}&=(1+x^2)^{-\frac12}(x\e{-i\vartheta}\ket{s_1}-\ket{s_2}),
\end{align}
\end{subequations}
 where $x\ne 0$ is real and $-\frac\pi 2\le\vartheta< \frac\pi 2$.

Using these definitions, Eq.~\eqref{Eq-OptDelta2Eq2} can be written as
\begin{equation}\label{Eq-OptDelta3Eq2}
 g_1 g_{23} \sin\vartheta=0 \mx{and} x g_1(x g_2+ g_{23}(x^2-1))\ge 0.
\end{equation}
Let us first discuss the case where $g_{23}\ne 0$.
In this situation we get\footnote{Remember, that we chose our basis such that 
 $g_{23}\ne 0$ implies $g_1\ne 0$.} from Eq.~\eqref{Eq-OptDelta3Eq2} the phase 
 $\vartheta= 0$.
The solutions of Eq.~\eqref{Eq-OptDelta2Eq1} are given by the real roots of the 
 polynomial of degree six in $x$,
\begin{equation}\label{e26465}
 x^2g_1^2(x^2g_{21}+g_{22}- 2 x g_{23})
  -(x g_2+ (x^2-1) g_{23})^2 (x^2g_{11}+g_{12})=0.
\end{equation}
(Since $g_{23}^2 g_{12}> 0$, this polynomial cannot be trivial.)

It now remains to consider the special case, where $g_{23}= 0$, but neither 
 $\ket\phi= \ket{s_1}$ nor $\ket\phi= \ket{s_2}$ is optimal.
If $g_{23}= 0$, then Eq.~\eqref{Eq-OptDelta2Eq1} reads
\begin{equation}
 x^2 (g_1^2 g_{21}- g_2^2 g_{11})= g_2^2 g_{12}- g_1^2 g_{22}.
\end{equation}
Assume that this equation has some solutions where $x$ is real (there might be 
 infinitely many).
None of these solutions leads to an optimal measurement, as we exclude by the 
 following argumentation.
Neither Eq.~\eqref{Eq-Eldar1x2e3} nor Eq.~\eqref{Eq-OptDelta2Eq2} depend on 
 $\vartheta$.
From the necessary and sufficient conditions for optimality (cf.\ end of 
 Sec.~\ref{s9110}), only the condition $\nu\equiv \bra{\phi^\perp} K^\dag 
 K\ket{\phi^\perp}<1$ remains to be satisfied.
Since Eq.~\eqref{e21583} does not depend on $\vartheta$, also $K$ is 
 independent of $\vartheta$.

Thus $\nu$ as a function of $\vartheta$ is of the form
\begin{equation}
 \nu(\vartheta)=(1+x^2)^{-1}
              \left(x^2 \kappa_{11}- 2\,x\Re(\e{-i\vartheta}\kappa_{12})+
                    \kappa_{22}\right),
\end{equation}
 where we defined $\kappa_{ij}= \bra{s_i}K^\dag K\ket{s_j}$.
In particular this function is continuous in $\vartheta$.
Assume, for a given $x\ne 0$ (satisfying Eq.~\eqref{Eq-Eldar1x2e3} and
 Eq.~\eqref{e25745}), there exists some $-\pi/2\le \vartheta<\pi/2$, such that 
 $\nu(\vartheta)<1$.
Then there exists an $\epsilon>0$, such that also $\vartheta+ \epsilon<\pi/2$ 
 and $\nu(\vartheta+\epsilon)<1$.
Hence for $\vartheta$ as well as $\vartheta+\epsilon$ all optimality conditions 
 would be satisfied.
But for both values, we get a different vector $\ket\phi$ and hence there would 
 be two different operators $E_?$, both being optimal.
This is a contradiction to Proposition~\ref{p14866}.
It follows that for $g_{23}= 0$, only $\ket\phi= \ket{s_1}$ and $\ket\phi= 
 \ket{s_2}$ can yield an optimal solution.
 
%..............................................................................%
\subsubsection{Summary for measurement type $(1,2)$}
Let us briefly summarize the algorithm to obtain the optimal measurement $E_?$ 
 for the case where $\rank E_1=1$ and $\rank E_2= 2$.

We construct some basis $(\ket{s_1},\ket{s_2})$ of $\supp\gamma_1$ as described 
 in the paragraph below Eq.~\eqref{Eq-OptDelta2Eq2}.
In a next step, we construct candidates for the basis 
 $(\ket\phi,\ket{\phi^\perp})$.
There are two cases:
\begin{itemize}
\item[(i)] $g_{23}=0$.
           If $g_{21}\ge g_{11}$, then $(\ket\phi= \ket{s_1}, 
            \ket{\phi^\perp}=\ket{s_2})$ is a candidate.
           If $g_{22}\ge g_{12}$, then $(\ket\phi= \ket{s_2},\ket{\phi^\perp}= 
            \ket{s_1})$ is a candidate.
\item[(ii)] $g_{23}\ne 0$.
            For any real root $x\ne 0$ of the polynomial in Eq.~\eqref{e26465}, 
              the basis $(\ket\phi,\ket{\phi^\perp})$ as defined in 
              Eq.~\eqref{e23160} is a candidate, where $\vartheta= 0$.
             A candidate in addition has to satisfy the second part of 
              Eq.~\eqref{Eq-OptDelta3Eq2} and Eq.~\eqref{Eq-Eldar1x2e3}.
\end{itemize}

For any of the candidate bases (if any), we construct $\sqrt \nu \ket n$ using 
 Eq.~\eqref{e31410}.
At most one of the bases will satisfy $\nu\equiv \sqrt\nu\bracket n n \sqrt \nu 
 <1$.
If such a basis exists, the optimal measurement is given by Eq.~\eqref{e18439}.

%------------------------------------------------------------------------------%
\subsection{The measurement class $[1,1]$}\label{s25151}
In Sec.~\ref{s12452} we have already seen, that if $\rank E_1+ \rank E_2= \rank 
 \gamma_1(= \rank \gamma_2)$, then both $E_1$ and $E_2$ are projectors.
Hence there are orthonormal bases $(\ket{\psi_1}, \ket{\psi_1^\perp})$ of 
 $\ker\gamma_2$ and $(\ket{\psi_2}, \ket{\psi_2^\perp})$ of $\ker\gamma_1$ such 
 that
\begin{equation}\label{e29323}
 E_1 =\pr{\psi_1} \mx{and} E_2 =\pr{\psi_2}.
\end{equation}
Since $\id- E_1- E_2\ge 0$, necessarily  $\bracket{\psi_1}{\psi_2}= 0$ must 
 hold.
Using this notation, Eq.~\eqref{e26860} is equivalent to
\begin{subequations}\label{e16014}\begin{gather}\label{e14050a}
 \abs{\bracket{\psi^\perp_1}{\psi_2}}^2 \bra{\psi_2} \gamma_2\ket{\psi_2} \ge
  \bra{\psi_1^\perp}\gamma_1 \ket{\psi_1^\perp}\\\label{e14050b}
 \abs{\bracket{\psi^\perp_2}{\psi_1}}^2 \bra{\psi_1} \gamma_1\ket{\psi_1} \ge
  \bra{\psi_2^\perp}\gamma_2 \ket{\psi_2^\perp}\\\label{Eq-Eldar2x2}
 \bracket{\psi_2^\perp}{\psi_1} \bra{\psi_1}\gamma_1 \ket{\psi_1^\perp}=
 \bracket{\psi_2}{\psi_1^\perp} \bra{\psi_2^\perp}\gamma_2 \ket{\psi_2},
\end{gather}\end{subequations}
 while Eq.~\eqref{e4736b} is satisfied identically.
(Note that these equations only follow if all vectors are normalized.)

%..............................................................................%
\subsubsection{Construction of a finite number of candidates for $E_?$}
\label{s9234}
Let $(\ket{k_{1i}})$ and  $(\ket{k_{2i}})$ be Jordan bases  (cf.\ 
 Appendix~\ref{a32712}) of $\ker\gamma_1$ and $\ker\gamma_2$, i.e., 
 $(\ket{k_{1i}})$ and $(\ket{k_{2i}})$ are orthonormal bases of $\ker\gamma_1$ 
 and $\ker\gamma_2$, respectively, such that $\bracket{k_{1l}}{k_{2l}}\ge 0$ 
 and for $i\ne j$ one has $\bracket{k_{1i}}{k_{2j}}= 0$.
Due to our assumptions in Eq.~\eqref{e6795} we have $0<\bracket{k_{1i}}{k_{2i}} 
 <1$.
We choose $\bracket{k_{11}}{k_{21}}\ge \bracket{k_{12}}{k_{22}}$.
In case of degenerate Jordan angles (i.e., $\bracket{k_{11}}{k_{21}}= 
 \bracket{k_{12}}{k_{22}}$), these bases are not unique
 and we then choose bases, such that $\bra{k_{21}}\gamma_1 \ket{k_{22}}= 0$.
We abbreviate
\begin{equation}\begin{split}
 g_{1i}= \bra{k_{2i}}\gamma_1 \ket{k_{2i}},&\quad
 g_{2i}= \bra{k_{1i}}\gamma_2 \ket{k_{1i}},\\
 g_{13}= \abs{\bra{k_{21}}\gamma_1 \ket{k_{22}}},&\quad
 g_{23}= \abs{\bra{k_{11}}\gamma_2 \ket{k_{12}}},\\
 g_\mu&= g_{\mu1}-g_{\mu2},
\end{split}\end{equation}
 and choose the global phases of the vectors $\ket{k_{12}}$ and $\ket{k_{22}}$ 
 in such a way that still $\bracket{k_{12}}{k_{22}}>0$ but also $g_{13} 
 \e{i\varphi}= \bra{k_{21}}\gamma_1 \ket{k_{22}}$ and $g_{23} \e{-i\varphi}= 
 \bra{k_{11}}\gamma_2 \ket{k_{12}}$, with $0\le \varphi< \pi$.
We let $\varphi= 0$ if $g_{13}=0$ or $g_{23}= 0$.
Finally we define $c= \bracket{k_{12}}{k_{22}}/ \bracket{k_{11}}{k_{21}}$.

After choosing the Jordan bases, let us first consider the case where 
 $\ket{\psi_1}= \ket{k_{21}}$.
Then one has (fixing the phases) $\ket{\psi_2}= \ket{k_{12}}$, 
 $\ket{\psi_1^\perp}= \ket{k_{22}}$, and $\ket{\psi_2^\perp}= \ket{k_{11}}$.
Eq.~\eqref{Eq-Eldar2x2} reduces to $c g_{23}= g_{13}$ and $\sin\varphi= 0$.
Analogously, in the case $\ket{\psi_1}= \ket{k_{22}}$, we obtain $c g_{13}= 
 g_{23}$ and $\sin\varphi= 0$.

We now consider the case, where $\ket{\psi_1}$ is not one of the basis vectors 
 $\ket{k_{2i}}$.
We fix the global phases of $\ket{\psi_\mu}$ and $\ket{\psi_\mu^\perp}$ and
 choose the parametrization
\begin{equation}\begin{split}\label{e30927}
 \ket{\psi_1}       &= (1+ x^2)^{-\frac12}
                        (\ket{k_{21}}+ x\e{i\vartheta}\ket{k_{22}}),\\
 \ket{\psi_1^\perp} &= (1+ x^2)^{-\frac12}
                        (x\e{-i\vartheta}\ket{k_{21}}-\ket{k_{22}}),\\
 \ket{\psi_2}       &= (1+ c^2x^2)^{-\frac12}
                        (-x \e{-i\vartheta}c\ket{k_{11}}+\ket{k_{12}}),\\
 \ket{\psi_2^\perp} &= (1+c^2x^2)^{-\frac12}
                        (-\ket{k_{11}}-x \e{i\vartheta}c\ket{k_{12}}),
\end{split}\end{equation}
 with the real parameters $x\ne 0$ and $-\pi/2\le \vartheta< \pi/2$.

Using these definitions, Eq.~\eqref{Eq-Eldar2x2} reads
\begin{multline}\label{Eq-Eldar1x1c}
 (x^2c^2+ 1)^2 \left(x g_1- \e{i(\vartheta+\varphi)}g_{13}+x^2 
  \e{-i(\vartheta+\varphi)}g_{13}\right)=\\
 c(x^2+1)^2 \left(x c g_2-\e{i(\vartheta-\varphi)}g_{23}+
  x^2c^2\e{-i(\vartheta-\varphi)}g_{23}\right)
\end{multline}
The imaginary part of this equation gives
\begin{multline}
 (x^2c^2+ 1)(1+ x^2)\\
 \cdot\left[
 (x^2c^2+ 1) \sin(\vartheta+\varphi) g_{13}-(1+ x^2) \sin(\vartheta-\varphi)c 
 g_{23}
 \right]= 0,
\end{multline}
 which can only hold if already the term in square brackets is zero.
This is the case if and only if
\begin{equation}\label{Eq-Eldar1x1Theta}
 A_1 \sin\vartheta = A_2 \cos\vartheta,
\end{equation}
 where
\begin{subequations}
\begin{align}
 A_1& =(c(x^2+1)g_{23}-(x^2c^2+1)g_{13})\cos \varphi,\\
 A_2& =(c(x^2+1)g_{23}+(x^2c^2+1)g_{13})\sin \varphi.
\end{align}
\end{subequations}
 
In order to get the solutions of Eq.~\eqref{Eq-Eldar1x1c}, we consider now its 
 real part,
\begin{multline}
(x^2c^2+1)^2 (x g_1+\cos(\vartheta+\varphi)g_{13}(x^2-1) )=\\
c(x^2+1)^2 (x c g_2+\cos (\vartheta-\varphi)g_{23}(x^2c^2-1)).
\end{multline}
Using the abbreviations
\begin{equation}\begin{split}
 B_1 &=(x^2c^2+1)^2x g_1-c(x^2+1)^2x c g_2,\\
 B_2 &=
 \left[(x^2c^2+1)^2g_{13}(x^2-1)-c(x^2+1)^2g_{23}(x^2c^2-1)\right]\cos\varphi,\\
 B_3 &=
 \left[(x^2c^2+1)^2g_{13}(x^2-1)+c(x^2+1)^2g_{23}(x^2c^2-1)\right]\sin\varphi,
\end{split}\end{equation}
 we get the equivalent expression
\begin{equation}\label{Eq-1x1Hx}
 B_1=B_3\sin\vartheta -B_2\cos\vartheta.
\end{equation}
Taking the square of this equation, multiplied by $(A_1^2+A_2^2)$, we obtain 
 due to Eq.~\eqref{Eq-Eldar1x1Theta} the polynomial (with a degree of at most 
 $8$ in $x^2$)
\begin{equation}\label{e31390}
B_1^2 (A_1^2+A_2^2)-(A_1B_2-A_2B_3)^2=0.
\end{equation}

This polynomial is trivial if and only if $g_{13}= g_{23}= 0$.
(In order to see this, we consider the highest and lowest order term, which 
 only can vanish, if $-(c g_{13}- g_{23})^2 \cos^2\varphi = (c g_{13}+ 
 g_{23})^2 \sin^2\varphi$ and if $-(c g_{23}- g_{13})^2 \cos^2\varphi = (c 
 g_{23}+ g_{13})^2 \sin^2\varphi$, respectively.
But due to our particular choice of the bases, this can only hold if already 
 $g_{13}= g_{23}= 0$.)
It is straightforward to see, that in this case none of the conditions in 
 Eq.~\eqref{e16014} depend on $\vartheta$.
Now suppose there is a solution of these conditions with $x\ne 0$.
Then any possible value of $\vartheta$ leads to a different, but optimal 
 measurement, in contradiction to Proposition~\ref{p14866}.

In any other situation we get from Eq.~\eqref{e31390} a finite number of real 
 solutions $x\ne 0$.
The corresponding value for $\vartheta$ can be obtained as follows.
If $A_1\ne 0$, then from Eq.~\eqref{Eq-Eldar1x1Theta} we have $\vartheta= 
 \arctan(A_2/A_1)$, while if $A_1= 0$ and $A_2\ne 0$, then $\vartheta= -\pi/2$.
If $A_1= A_2= 0$, then $\sin\varphi= 0$ and
\begin{equation}
 x^2=\frac1c\frac{cg_{23}-g_{13}}{cg_{13}-g_{23}},
\end{equation}
 where from $x^2>0$ it follows that $(cg_{23}-g_{13})(cg_{13}-g_{23})> 0$.
From Eq.~\eqref{Eq-1x1Hx} we have
\begin{equation}\label{e1890}
 2 \cos\vartheta\, g_{13} g_{23} (cg_{23}-g_{13})=
   x c (g_{23}^2 g_1-g_{13}^2 g_2),
\end{equation}
 which can be used in order to obtain $\vartheta$.

%..............................................................................%
\subsubsection{Summary for measurement class $[1,1]$}
In a first step we construct Jordan bases as described in the first paragraph 
 of Sec.~\ref{s9234}.
Then we collect candidates for the bases $(\ket{\psi_i},\ket{\psi_i^\perp})$:
\begin{itemize}
\item[(i)] If $\sin\varphi= 0$ and  $c g_{23}= g_{13}$, then $(\ket{\psi_1}= 
 \ket{k_{21}}, \ket{\psi_1^\perp}=\ket{k_{22}})$ and $(\ket{\psi_2}= 
 \ket{k_{12}}, \ket{\psi_2^\perp}= \ket{k_{11}})$ is a candidate.
\item[(ii)] If $\sin\varphi= 0$ and $c g_{13}= g_{23}$, then $(\ket{\psi_1}= 
 \ket{k_{22}}, \ket{\psi_1^\perp}=\ket{k_{21}})$ and $(\ket{\psi_2}= 
 \ket{k_{11}}, \ket{\psi_2^\perp}= \ket{k_{12}})$ is a candidate.
\item[(iii)] For any real root $x\ne 0$ of Eq.~\eqref{e31390}, we get a unique 
 value for $\vartheta$ from Eq.~\eqref{Eq-Eldar1x1Theta} (if $A_1\ne 0$ or 
 $A_2\ne 0$) or from Eq.~\eqref{e1890} (if $A_1= 0$ and $A_2= 0$).
If $A_1\ne 0$ or $A_2\ne 0$, then in addition Eq.~\eqref{Eq-1x1Hx} has to 
 hold.
For each $(x, \vartheta)$, we obtain the candidates from Eq.~\eqref{e30927}.
\end{itemize}

At most one of the candidates will satisfy Eq.~\eqref{e14050a} and 
 Eq.~\eqref{e14050b}.
If such a candidate exists, the optimal measurement is provided by 
 Eq.~\eqref{e29323}.

%------------------------------------------------------------------------------%
\subsection{Examples}\label{s2290}
\begin{figure}\center
\includegraphics[width=.75\textwidth]{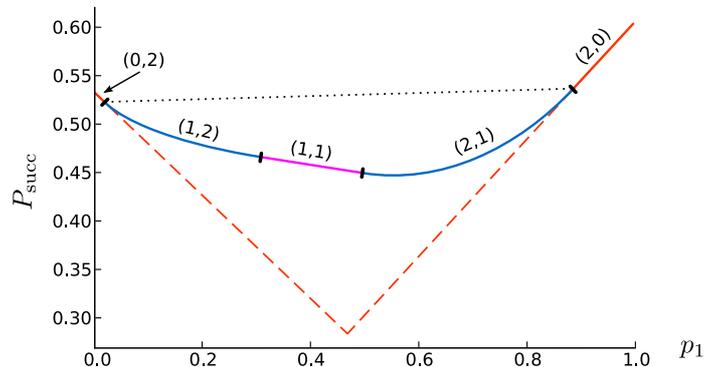}
\caption{\label{Fig-Example1}
Optimal success probability of the states given in Eq.~\eqref{Eq-Example1}, 
 depending on the relative probability $p_1$ of the occurrence of $\rho_1$ 
 (solid line).
The dashed lines denote the success probability of single state detection 
 (lower bound) and the dotted line corresponds to a simple upper bound.
In brackets we denote the measurement types as defined in Sec.~\ref{s12452}.
}
\end{figure}

We want to discuss a few examples of USD in four dimensions, which demonstrate 
 the structure of the previous results.
We consider three examples which belong to case [vii] in the flowchart in 
 Fig.~\ref{Fig-USDFlowChart}.
Thus these pairs of states in particular are strictly skew and do not posses a 
 common block diagonal structure.
In Fig.~\ref{Fig-Example1}
the optimal success probability of the two states
\begin{equation}\label{Eq-Example1}
\rho_{1}  =\frac{1}{3}
\begin{pmatrix}
1 & 0 & 0 & 0 \\
0 & 2 & 0 & 0 \\
0 & 0 & 0 & 0 \\
0 & 0 & 0 & 0
\end{pmatrix} \mx{and}
\rho_{2}  =\frac{1}{45}
\begin{pmatrix}
11 & 10 & 12 & 10 \\
10 & 10 & 10 & 10 \\
12 & 10 & 14 & 10 \\
10 & 10 & 10 & 10
\end{pmatrix},
\end{equation}
 is given in dependence of the \apriori\ probability of $p_1$ of $\rho_1$ 
 (solid line).
Following the results of Sec.~\ref{s194} and Sec.~\ref{s3330} we can directly 
 calculate the probability range where single state detection (class $[0,2]$) 
 and the fidelity form measurement are optimal (roughly class $[2,2]$).
The remaining optimal measurements belong to the classes $[1,2]$ or $[1,1]$ 
 which are obtained by following Sec.~\ref{Sec-DeltaOneDim} or 
 Sec.~\ref{s25151}, respectively.
We also plotted the ``bound triangle'' which can be easily calculated for any 
 pair of density operators.
The lower bounds correspond to single state detection (dashed lines) and the 
 upper bound connects the points where single state detection stops being 
 optimal (dotted line).
The latter one is an upper bound due to the convexity of the success 
 probability function $P_\mathrm{succ}(p_1)$.

\begin{figure}\center
\begin{minipage}{.75\textwidth}\center
\includegraphics[width=\textwidth]{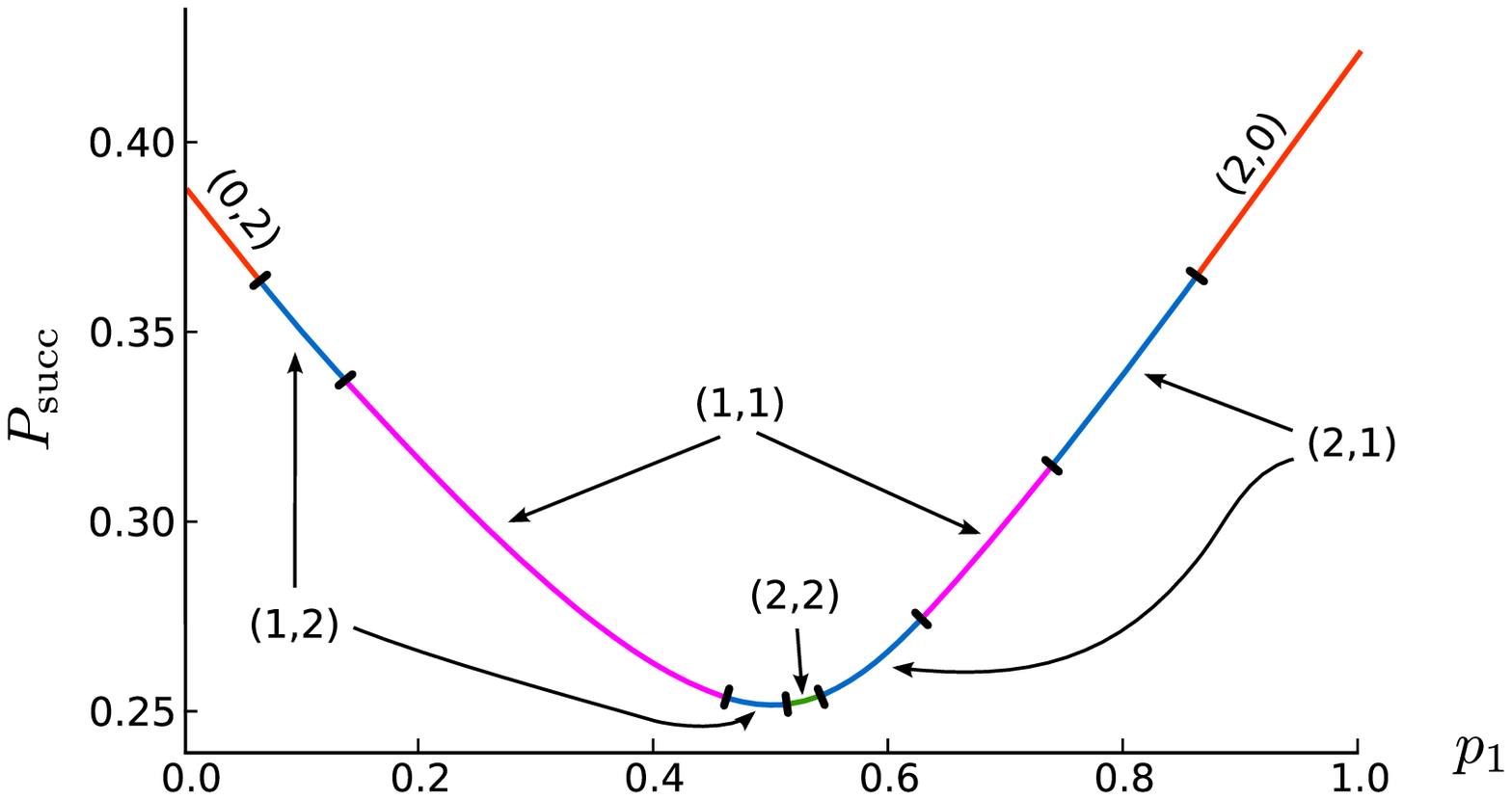}
\end{minipage}\hspace{3em}(A)
\begin{minipage}{.75\textwidth}\center
\includegraphics[width=\textwidth]{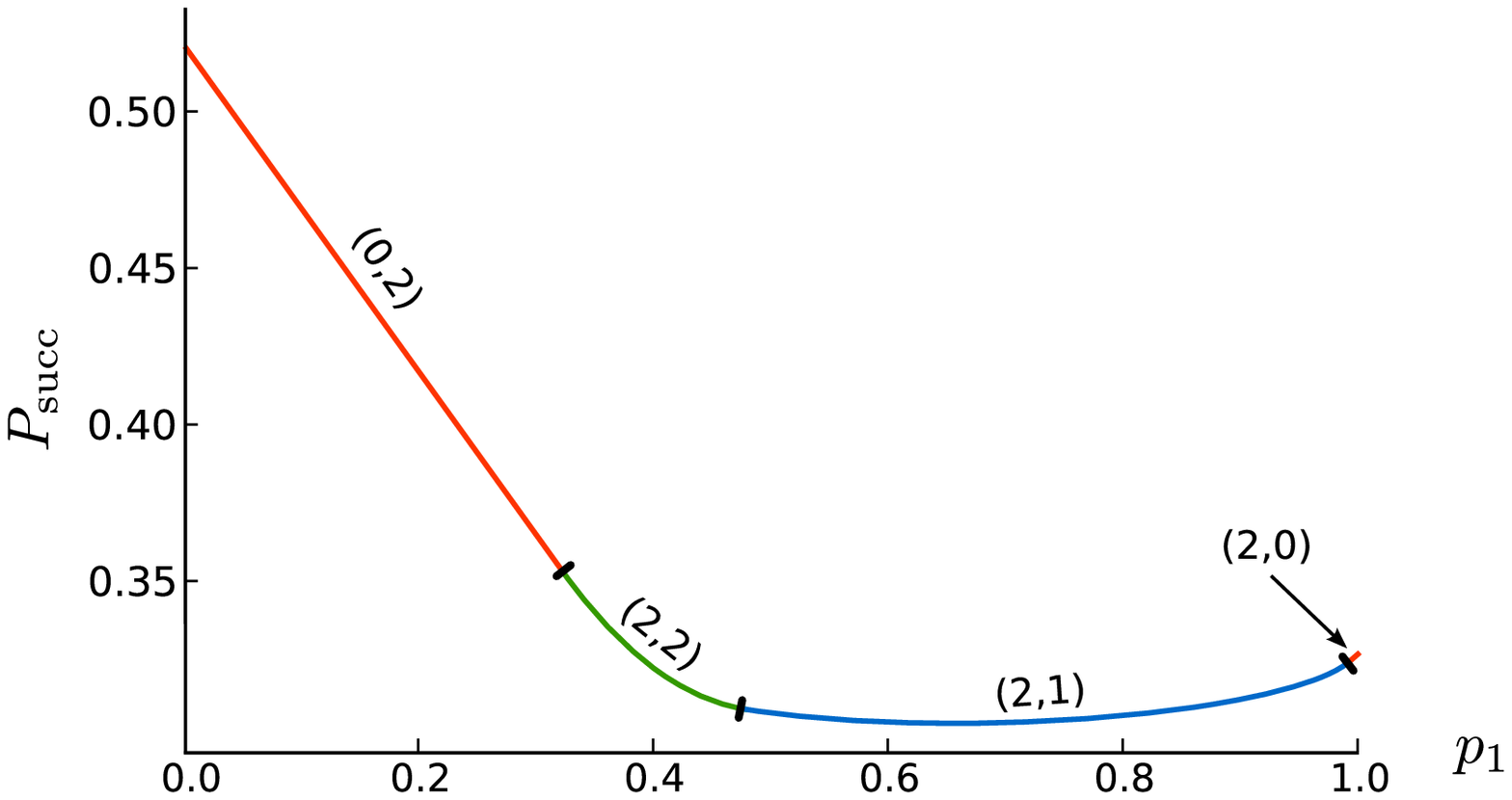}
\end{minipage}\hspace{3em}(B)
\caption{\label{Fig-Examples2}
Optimal success probability depending on the relative probability $p_1$ of the 
 occurrence of $\rho_1'$.
(A): A ``random'' example.
(B): An asymmetric example, cf.\ Eq.~\eqref{Eq-Examples2}.
In brackets we denote the measurement types as defined in Sec.~\ref{s12452}.
For details see text in Sec.~\ref{s2290}.
}
\end{figure}
The next example (Fig.~\ref{Fig-Examples2} (A)) was found by generating 
 ``randomly'' pairs of density operators.
The data of the states is available as supplemental material.
This example shows that the measurement types $(1,1)$, $(1,2)$, and $(2,1)$ can 
 appear in more than one probability range.
This is not possible for the other measurement types.
It is also an example for a pair of states, where all possible measurement 
 types can be optimal in some probability range.

The last example (Fig.~\ref{Fig-Examples2} (B)), given by
\begin{equation}\label{Eq-Examples2}
\begin{split}
\rho_1' & =\left(\frac{1}{2}+\sqrt{\frac{5}{22}}\right)
\pr{0}+\left(\frac{1}{2}-\sqrt{\frac{5}{22}}\right)\pr{1},\\
\rho_2' & =\frac{5}{46}\pr{v}+\frac{41}{46}\pr{w}, \quad\text{with}\\
\ket{w} &=\frac{1}{2\sqrt{41}}\left((-1)^{1/7}
 \left((\sqrt{22}+2\sqrt 5)\ket{0}+(\sqrt{22}-2\sqrt 5)\ket{1}\right)+
 2\sqrt{10}\left(\ket{2}+\ket{3}\right)\right),\\
\ket{v} &=\frac{1}{\sqrt{10}}\left(((-1)^{4/21})^*\ket{0}
 +(-1)^{17/21}\ket{1}+2\sqrt 2 (-1)^{1/5}\ket{2}\right),
\end{split}
\end{equation}
 is devoted to show that there is no deeper general structure which optimal 
 measurement classes can be connected with each other.
Here, the fidelity form measurement directly follows the single state detection 
 measurement.
This example also shows a high asymmetry in the success probability function 
 $P_\mathrm{succ}(p_1)$.

% vim:ft=tex ==================================================================%
\section{General strategy}\label{s23733}
\begin{figure}\center
\includegraphics[width=0.80\textwidth]{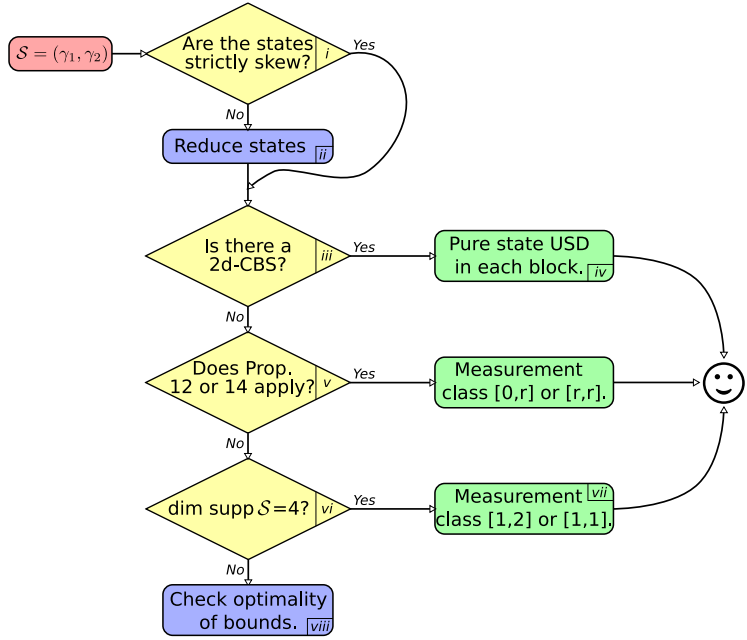}
\caption{\label{Fig-USDFlowChart}
Flow chart of a generic strategy to solve/reduce the USD of two mixed states.
For details see text in Sec.~\ref{s23733}.
}
\end{figure}

In this section we want to summarize the known results for the unambiguous 
 discrimination of two mixed states by suggesting a strategy in order to find 
 the optimal success probability, cf.\ Fig.~\ref{Fig-USDFlowChart}.

In step [i] we check whether the pair of weighted density operators $\mathcal 
 S=(\gamma_1, \gamma_2)$ is \emph{strictly skew}.
A pair $\mathcal S$ is called strictly skew, if $\supp \gamma_1\cap 
 \supp\gamma_2= \{0\}$, $\supp\gamma_1\cap \ker\gamma_2= \{0\}$, and 
 $\ker\gamma_1\cap \supp\gamma_2= \{0\}$.
If the pair fails to be strictly skew, one can use the reduction method 
 described in the discussion of Lemma~\ref{l2980} (Fig.~\ref{Fig-USDFlowChart} 
 [ii]).
After this reduction, it suffices to find the optimal measurement $\mathcal E'$ 
 of the strictly skew pair $\mathcal S'$.

From now on we assume that the states are strictly skew.
In the next step (Fig.~\ref{Fig-USDFlowChart} [iii]) we check with the 
 commutator criteria shown in Ref.~\cite{Kleinmann:2007JPA} whether the two 
 states have an at most two-dimensional common block diagonal structure.
If this is the case in Ref.~\cite{Kleinmann:2007JPA} it was shown how to 
 construct diagonalizing Jordan bases, which then can be used to compose the 
 optimal measurement from the pure state case (Fig.~\ref{Fig-USDFlowChart} 
 [iv]), as shown in Ref.~\cite{Bergou:2006PRA,Kleinmann:2007JPA}.

For states without an at most two-dimensional common block diagonal structure, 
 optimality of one of the two generally solved measurement classes, i.e., 
 single state detection (Proposition~\ref{p6402}) or the fidelity form 
 measurement (Proposition~\ref{p22024}), can be checked 
 (Fig.~\ref{Fig-USDFlowChart} [v]).
Otherwise the optimal measurement can be calculated if the two states 
 effectively act on a four-dimensional Hilbert-space ($\dim\supp \mathcal S 
 =4$).
Here the optimal measurement of the two remaining measurement classes $[1,2]$ 
 and $[1,1]$ can be calculated and checked according to 
 Sec.~\ref{Sec-DeltaOneDim} and Sec.~\ref{s25151}, respectively 
 (Fig.~\ref{Fig-USDFlowChart} [vi]).
In principle one could also find optimal measurements of states which have a 
 four-dimensional common block diagonal structure, because then blockwise 
 combination of optimal measurements yields the optimal measurement for the 
 complete states.
Unfortunately, there is so far no known constructive method to identify such 
 blocks.
This question is left for further investigations.

The last possibility is to check optimality of upper and lower bounds on the 
 optimal success probability (Fig.~\ref{Fig-USDFlowChart} [viii]).
Examples of such bounds were presented e.g.\ in 
 Ref.~\cite{Rudolph:2003PRA,Herzog:2005PRA,Raynal:2005PRA,Zhou:2007PRA}.
For some of these bounds, optimality conditions are known, while in the 
 remaining cases one can use Corollary~\ref{c24797} in order to check for 
 optimality.

If the procedure sketched here fails to deliver the optimal solution then there 
 is up to now no systematic method known to find an analytic expression for the 
 optimal USD measurement.

% vim:ft=tex ==================================================================%
\section{Conclusions}\label{s23013}
We analyzed the unambiguous discrimination of two mixed states $\rho_1$ and 
 $\rho_2$ and the properties of an optimal measurement strategy.
We first showed (cf.\ Proposition~\ref{p6437}) that any unambiguous state 
 discrimination measurement is completely determined by the measurement 
 operator $E_?$, the operator which corresponds to the inconclusive measurement 
 result.
A further analysis for optimal measurements showed (cf.\ Theorem~\ref{t4134}) 
 that the rank of $E_?$ is determined by the structure of the input states, 
 $\rank E_?= \rank\rho_1\rho_2$.

This fact leads to one of our main results, namely, that the optimal 
 measurement for a given pair of mixed states and given \apriori\ probabilities 
 is unique (cf.\ Proposition~\ref{p14866}).
This uniqueness might have interesting consequences, e.g.\ in the analysis of 
 the ``complexity'' of the optimal measurement.
Interesting questions here are whether the optimal measurement is von-Neumann 
 or whether the optimal measurement is non-local, as e.g.\ discussed in
 Ref.~\cite{Kleinmann:2005PRA}.

Eldar \etal\ in Ref.~\cite{Eldar:2004PRA} provided necessary and sufficient 
 conditions for a measurement to be optimal, but in many situations this result 
 was not operational.
We simplified these optimality conditions in Corollary~\ref{c24797}, which now 
 can be applied directly to a given measurement.

As an application of this result, we analyzed the single state detection case, 
 where the measurement only detects one of the two states.
Although this measurement may seem to be pathological, it turns out to be 
 always optimal for a finite region of the \apriori\ probabilities of the 
 states.
We derive an analytical expression for the bounds of this region.

Finally, we constructed the optimal measurement for the unambiguous 
 discrimination of two mixed states having $\rank\rho_1= 2= \rank\rho_2$ (due 
 to a results by Raynal \etal\ in Ref.~\cite{Raynal:2003PRA}, this construction 
 can be extended to the case where one of the density operators has rank 2 and 
 the other state has arbitrary rank).
The solution splits into 6 different types, each of them requiring a different 
 treatment.
This solution is analytical, but in certain cases the roots of a high order (up 
 to degree 8) polynomial are needed.
Due to the complicated structure of this solution it may turn out to be quite 
 difficult to analyze the next step, namely the discrimination of two mixed 
 states with $\rank\rho_1= 3= \rank\rho_2$.

It would be interesting to find strategies in order to detect symmetries and 
 solvable substructures in unambiguous state discrimination, e.g.\ to find 
 four-dimensional common block diagonal structures, analogously to the result 
 for two-dimensional common block diagonal structures in 
 Ref.~\cite{Kleinmann:2007JPA}.
Also, there are only a few results on the optimal unambiguous discrimination on 
 more than two states 
 \cite{Chefles:1998PLA,Chefles:1998PLA2,Peres:1998JPA,Zhang:2006PLA,%
 Eldar:2004PRA,Samsonov:2009PRA}.
We think that several concepts presented in this contribution may also 
 generalize to the discrimination of many states.

\begin{acknowledgement}
We would like to thank J.~Bergou, T.~Meyer, Ph.~Raynal, Z.~Shadman, and 
 R.~Unanyan for valuable discussions.
This work was partially supported by the EU Integrated Projects SECOQC, SCALA, 
 OLAQUI, QICS and by the FWF.
\end{acknowledgement}

\appendix
% vim:ft=tex ==================================================================%
\section{Appendix: Technical statements}\label{s26828}
We put the following two Lemmata for completeness, but without a proof.

\begin{lemma}
Let $\mathscr A$, $\mathscr B$ and $\mathscr C$ be subspaces of $\Hilbert$ with 
 $\mathscr A \perp \mathscr C$ and $\mathscr B \perp \mathscr C$.
Then $\mathscr A \cap (\mathscr B+ \mathscr C)= \mathscr A\cap \mathscr B$.
\end{lemma}

\begin{lemma}\label{l12912}
Let $A$ and $B$ be operators on $\Hilbert$ with $B\Hilbert\cap \ker A= \{0\}$.
Then $\ker AB= \ker B$.
\end{lemma}

The useful Eq.~\eqref{e25118} is based on the following
\begin{lemma}\label{l2502}
Let $X$ be an operator and $(\Pi_k)$ be a family of projectors with 
 $\ker(\sum_k\Pi_k)= \{0\}$.
For $k\ne l$, assume that $\Pi_k X \Pi_l= 0$.
(Note that $\Pi_k\Pi_l\ne 0$ is allowed.).
Then $\ker X= \sum_k (\ker X \cap \Pi_k\Hilbert)$.
\end{lemma}
\begin{proof}
The ``$\supset$'' part is obvious.
For the contrary let $\ket\Phi\in \ker X$.
Then there exist vectors $\ket{\varphi_k}\in \Pi_k\Hilbert$ such that 
 $\ket\Phi= \sum_k \ket{\varphi_k}$.
We have
\begin{equation}
\sum_l\Pi_l X \ket{\varphi_k}= \Pi_k X \ket{\varphi_k}=\Pi_k X \ket\Phi= 0.
\end{equation}
Since $\sum_k\Pi_k$ has a trivial kernel, it follows that $X\ket{\varphi_k}= 
 0$, i.e., $\ket{\varphi_k}\in \ker X$.
\end{proof}

The following Lemma lists properties of non-orthogonal projectors, i.e., 
 idempotent operators which are not self-adjoint.
Such operators where considered e.g.\ in 
 Ref.~\cite{Afriat:1957PCPS,Greville:1974SJAM}.
Each of the statements (ii) -- (iv) is a valid definition for the oblique 
 projector, where (ii) is the formal definition, (iii) is an explicit 
 construction and (iv) are the central properties for our purposes.
\begin{lemma}\label{l10547}
Let $\Lambda$ and $\Pi$ be two (self-adjoint) projectors on $\Hilbert$ with 
 $\Lambda\Hilbert\cap (\Pi\Hilbert)^\perp= \{0\}$ and 
 $(\Lambda\Hilbert)^\perp\cap \Pi\Hilbert= \{0\}$.
For an operator $Q$ the following statements are equivalent:
\begin{itemize}
\item[(i)]   $Q$ is the (unique) \emph{oblique projector} from 
              $\Lambda\Hilbert$ to $\Pi\Hilbert$.
\item[(ii)]  $Q^2=Q$, $Q\Hilbert= \Pi\Hilbert$, and $\ker Q= \ker \Lambda$.
\item[(iii)] $Q$ is the Moore-Penrose inverse of $\Lambda\Pi$.
\item[(iv)]  $Q\Lambda= Q$, $\Pi Q= Q$, $\Lambda Q=\Lambda$ and $Q\Pi= \Pi$.
\end{itemize}
\end{lemma}
\begin{proof}
(ii) formally defines (i).
In order to show that (iii) follows from (ii), we write $Q$ in its singular 
 value decomposition, $Q= \sum q_i \ketbra{\pi_i}{\lambda_i}$ with $q_i>0$ and 
 $(\ket{\pi_i})$ and $(\ket{\lambda_i})$ an appropriate pair of orthonormal and 
 complete bases of $\Pi\Hilbert$ and $\Lambda\Hilbert$, respectively.
Then $Q^2= Q$ is equivalent to $\bracket{\lambda_k}{\pi_k}= q_k^{-1}$ and 
 $\bracket{\lambda_i}{\pi_j}= 0$ for $i\ne j$.
Since $\Lambda\Pi= \sum_i q_i^{-1} \ketbra{\lambda_i}{\pi_i}$, it immediately 
 follows that $Q$ is the Moore-Penrose inverse of $\Lambda\Pi$.

From the explicit form of the Moore-Penrose inverse, $Q= \sum_i 
 \bracket{\lambda_i}{\pi_i}^{-1} \ketbra{\pi_i}{\lambda_i}$, one easily 
 verifies the properties in (iv).

We have from (iv) that $Q\Hilbert= \Pi Q\Hilbert \subset \Pi\Hilbert= 
 Q\Pi\Hilbert \subset Q \Hilbert$ and hence $Q\Hilbert= \Pi\Hilbert$ and 
 analogously $\ker Q= (Q^\dag\Hilbert)^\perp= (\Lambda\Hilbert)^\perp= 
 \ker\Lambda$.
Finally, $Q^2= Q(\Pi Q)= (Q\Pi)Q= \Pi Q= Q$.
\end{proof}

%==============================================================================%
\section{Appendix: Proof of Corollary \ref{c24797}}\label{s18521}
%------------------------------------------------------------------------------%
\subsection{Necessity}
Let us abbreviate $\bar\mu= 3-\mu$, $Z_\mu= \Lambda_\mu Z \Lambda_\mu$ and 
 $Y_{\bar\mu}= \Lambda_\mu Z \Lambda_{\bar\mu}$.
From Theorem~\ref{t18481} it follows that for any optimal measurement we have 
 ($\mu= 1,2$)
\begin{subequations}
\begin{gather}\label{e19542a}
 Z_\mu- \Lambda_\mu \gamma_\mu \Lambda_\mu \ge 0,\\\label{e19542b}
 (Z_\mu- \Lambda_\mu \gamma_\mu \Lambda_\mu) E_\mu= 0,
\end{gather}
\end{subequations}
 and
\begin{subequations}\begin{align}
 Z_\mu (\Lambda_\mu -E_\mu) &=  Y_{\bar\mu} E_{\bar\mu} \Lambda_\mu,
  \label{e9392a}\\
 Y_{\bar\mu} (\Lambda_{\bar\mu} -E_{\bar\mu}) &= Z_\mu E_\mu \Lambda_{\bar\mu},
  \label{e9392b}
\end{align}\end{subequations}
 where the last two equations follow from $\Lambda_\mu ZE_? \Lambda_\nu= 0$ 
 with $E_?=\id- E_\mu- E_{\bar\mu}$.
From $Z\ge 0$ we have $Z_\mu \ge 0$ and $Y_{\bar\mu}= Y^\dag_\mu$.
We find
\begin{equation}\label{e18392}\begin{aligned}
 Z_\mu - \Lambda_\mu \gamma_\mu \Lambda_\mu &= (\Lambda_\mu- E_\mu)(Z_\mu 
  -\gamma_\mu) (\Lambda_\mu- E_\mu) \\
  &= \Lambda_\mu E_{\bar\mu} Y_{\bar\mu}^\dag (\Lambda_\mu - E_\mu)-
          (\Lambda_\mu- E_\mu) \gamma_\mu (\Lambda_\mu- E_\mu) \\
  &= \Lambda_\mu E_{\bar\mu} \gamma_{\bar\mu} E_{\bar\mu} \Lambda_\mu -
          (\Lambda_\mu- E_\mu) \gamma_\mu (\Lambda_\mu- E_\mu) \\
  &= \Lambda_\mu E_? \gamma_{\bar\mu} E_? \Lambda_\mu -
     \Lambda_\mu E_? \gamma_\mu E_? \Lambda_\mu.
\end{aligned}\end{equation}
Thus Eq.~\eqref{e26860a}, Eq.~\eqref{e26860b} follow from Eq.~\eqref{e19542a}.
From Eq.~\eqref{e19542b} we have
\begin{equation}\label{e25448}
 \Lambda_\mu E_?(\gamma_{\bar\mu}- \gamma_\mu) E_? E_\mu= 0.
\end{equation}

Combining Eq.~\eqref{e9392a} and Eq.~\eqref{e9392b} we obtain
\begin{equation}
 Y_{\bar\mu}\Lambda_\mu = Z_\mu E_\mu \Lambda_{\bar\mu}\Lambda_\mu +
 Z_\mu(\Lambda_\mu- E_\mu)
\end{equation}
and hence from Eq.~\eqref{e9392a} and due to Eq.~\eqref{e19542b},
\begin{equation}
\begin{aligned}
 \Lambda_{\bar\mu} Z_\mu(\Lambda_\mu- E_\mu)&= (Y_\mu\Lambda_{\bar\mu})^\dag
   E_{\bar\mu}\Lambda_\mu\\
  &= \Lambda_{\bar\mu}
  \Lambda_\mu E_{\bar\mu} \gamma_{\bar\mu} E_{\bar\mu} \Lambda_\mu+ 
   (\Lambda_{\bar\mu}- E_{\bar\mu}) \gamma_{\bar\mu} E_{\bar\mu} \Lambda_\mu\\
  &= \Lambda_{\bar\mu}(\Lambda_\mu- E_\mu)Z_\mu(\Lambda_\mu-E_\mu)+
   (\Lambda_{\bar\mu}- E_{\bar\mu}) \gamma_{\bar\mu} E_{\bar\mu} \Lambda_\mu,
\end{aligned}
\end{equation}
where in the last step we used the result from Eq.~\eqref{e18392}.
Thus we have found $\Lambda_{\bar\mu} E_\mu \gamma_\mu (\Lambda_\mu- E_\mu)= 
 (\Lambda_{\bar\mu}- E_{\bar\mu}) \gamma_{\bar\mu} E_{\bar\mu} \Lambda_\mu$, 
 i.e., Eq.~\eqref{e26860c} follows.

This equation together with Eq.~\eqref{e25448} for $\mu= 1$ and $\mu= 2$, 
 finally yields Eq.~\eqref{e4736b}.

%------------------------------------------------------------------------------%
\subsection{Sufficiency}
We first get rid of the non-skew parts of $\mathcal S$:
\begin{lemma}\label{l30817}
With the definitions and preliminaries as in Proposition~\ref{p2770}, if $E_?$ 
 is a proper USD measurement of $\mathcal S$ and satisfies Eq.~\eqref{e4736} 
 for $\mathcal S$, then $E_?^\mathrm{skew}=E_?+ (\id- \Pi_\mathrm{skew})$ is a 
 proper USD measurement of $\mathcal S^\mathrm{skew}$ and satisfies 
 Eq.~\eqref{e4736} for $\mathcal S^\mathrm{skew}$.
\end{lemma}
Hence, if we further show, that from the second part of the Lemma, it follows 
 that $E_?^\mathrm{skew}$ is optimal for $\mathcal S^\mathrm{skew}$, then we 
 have due to Proposition~\ref{p2770}, that also $E_?$ is optimal for $\mathcal 
 S$.
\begin{proof}[Proof of Lemma~\ref{l30817}.]
We denote by $\Sigma_1$ the projector onto $\supp\gamma_1\cap \ker\gamma_2$, by 
 $\Sigma_2$ the projector onto $\supp\gamma_2\cap \ker\gamma_1$ and by 
 $\Pi_\mu$ the projector onto $\supp\gamma_\mu$.
Then multiplication of Eq.~\eqref{e4736a} by $\Sigma_1$ yields
\begin{equation}
 \Sigma_1 E_? \gamma_2 E_? \Sigma_1 - \Sigma_1 E_? \gamma_1 E_? \Sigma_1 \ge 0.
\end{equation}
Since for a USD measurement $\Pi_1 E_? \gamma_2= \Pi_1 \gamma_2$, we have 
 $\Sigma_1\gamma_2= 0$ and only the second term remains, which henceforth must 
 vanish.
This yields $\sqrt{\gamma_1} E_?  \Sigma_1= 0$ or equivalently $\Pi_1 
 E_?\Sigma_1= 0$.
A further multiplication from the left by $\Sigma_1$ together with the property 
 $E_?\ge 0$ proves that $E_? \Sigma_1= 0$.
An analogous argument can be used in order to show $E_? \Sigma_2= 0$.
Now, following the same lines of argument as in the proof of 
 Proposition~\ref{p2770}, it follows that $E_?^\mathrm{skew}$ is a proper USD 
 measurement for $\mathcal S^\mathrm{skew}$.

From $E_?\Sigma_\mu= 0$ it in particular follows that $E_?(\gamma_2- 
 \gamma_1)E_?=  E_?^\mathrm{skew}\Pi_\mathrm{skew}(\gamma_2- 
 \gamma_1)\Pi_\mathrm{skew} E_?^\mathrm{skew}$.
Using
\begin{equation}
 \ker\Pi_\mathrm{skew}\gamma_\mu\Pi_\mathrm{skew}=
  \ker \gamma_\mu(\id- \Sigma_\mu)= \Sigma_\mu\Hilbert\oplus \ker\gamma_\mu,
\end{equation}
 it is now straightforward to show that $E_?^\mathrm{skew}$ satisfies 
 Eq.~\eqref{e4736} for $\mathcal S^\mathrm{skew}$.
\end{proof}

It remains to consider the case where $\supp\gamma_1\cap \ker\gamma_2= \{0\}= 
 \supp\gamma_2\cap \ker\gamma_1$.
We define $R_\mu$ to be the oblique projector from $\ker\gamma_{\bar\mu}$ to 
 $\ker\gamma_\mu$.
Note that $R_\mu= R_{\bar\mu}^\dag$.
We furthermore denote by $Q_\mu$ the oblique projector from 
 $\ker\gamma_{\bar\mu}\cap\supp \mathcal S$ to
 $\supp\gamma_\mu\cap (\ker\gamma_1+ \ker\gamma_2)$.
Then the multiplication of Eq.~\eqref{e4736b} by $Q_\mu^\dag$ from the left and 
 by $Q_\mu$ from the right yields
\begin{equation}\label{e9323}
 \Lambda_\mu E_? (\gamma_{\bar\mu}- \gamma_\mu) E_? E_\mu= 0.
\end{equation}

Let us define
\begin{align}
 V_\mu &= \Lambda_\mu E_?(\gamma_{\bar\mu}- \gamma_\mu)E_? \Lambda_\mu
          +\Lambda_\mu \gamma_\mu \Lambda_\mu, \\\label{e4901}
 W_{\bar\mu} &= (R_{\bar\mu}(\Lambda_{\bar\mu} - E_{\bar\mu}) + \Lambda_\mu 
                 E_{\bar\mu} ) V_{\bar\mu}.
\end{align}
Then, using Eq.~\eqref{e26860c},\eqref{e9323}, we have
\begin{equation}\label{e12457}
\begin{aligned}
 V_\mu(\Lambda_\mu- E_\mu)&= \Lambda_\mu E_?(\gamma_{\bar\mu}- \gamma_\mu)E_?
 \Lambda_\mu+ \Lambda_\mu\gamma_\mu E_? \Lambda_\mu \\&=
 \Lambda_\mu E_? \gamma_{\bar\mu} E_? \Lambda_\mu+
 E_\mu \gamma_\mu E_? \Lambda_\mu \\&=
 -\Lambda_\mu E_{\bar\mu} \gamma_{\bar\mu} E_? \Lambda_\mu- 
 R_{\bar\mu}E_?\gamma_\mu E_?  \Lambda_\mu \\&=
 -(\Lambda_\mu E_{\bar\mu} + R_{\bar\mu}E_?) \gamma_{\bar\mu} E_?  \Lambda_\mu
  \\&=
 (\Lambda_\mu E_{\bar\mu} + R_{\bar\mu}(\Lambda_{\bar\mu}- E_{\bar\mu}))
  \gamma_{\bar\mu} E_{\bar\mu} \Lambda_\mu
 \\&= W_{\bar\mu} E_{\bar\mu} \Lambda_\mu.
\end{aligned}
\end{equation}
And a similar equation holds for $W_{\bar\mu}$:
\begin{equation}\label{e3370}
\begin{aligned}
 W_{\bar\mu}(\Lambda_{\bar\mu}- E_{\bar\mu})=&
 R_{\bar\mu} E_?(\gamma_\mu- \gamma_{\bar\mu})E_? \Lambda_{\bar\mu}+
 R_{\bar\mu} (\Lambda_{\bar\mu}- E_{\bar\mu})\gamma_{\bar\mu}
             (\Lambda_{\bar\mu}- E_{\bar\mu})\\&\quad +\Lambda_\mu 
              E_{\bar\mu}\gamma_{\bar\mu}(\Lambda_{\bar\mu}- E_{\bar\mu})\\=&
 R_{\bar\mu} E_? \gamma_\mu E_? \Lambda_{\bar\mu}-
  \Lambda_\mu E_?\gamma_\mu E_? \Lambda_{\bar\mu} \\=&
  -(-E_\mu-(\Lambda_\mu-E_\mu))\gamma_\mu E_\mu \Lambda_{\bar\mu} \\=&
  V_\mu E_\mu \Lambda_{\bar\mu}.
\end{aligned}
\end{equation}
We combine these two equations and find $W_{\bar\mu}= V_\mu (E_\mu 
 \Lambda_{\bar\mu}+ (\Lambda_\mu- E_\mu)R_{\bar\mu})$, i.e., in comparison with 
 the definition in Eq.~\eqref{e4901}, $W_{\bar\mu}= W_\mu^\dag$.
Furthermore, we have $W_1 V_1^- V_1= W_1$ and $V_2= W_1 V_1^- W_1^\dag$, where 
 $V_1^-$ denotes the inverse of $V_1$ on its support.
(The second identity follows from $W_1 V_1^- W_1^\dag= W_1(\Lambda_1- E_1)R_2+ 
 W_1 E_1 \Lambda_2$.)

We construct the operator $Z$ as $Z= T V_1 T^\dag$ with $T=Q_1+ Q_2 W_1 V_1^-$.
Since $V_1\ge 0$ we have $Z\ge 0$.
From our previous considerations, $\Lambda_\mu Z \Lambda_\mu= V_\mu$ and 
 $\Lambda_\mu Z \Lambda_{\bar\mu}= W_{\bar\mu}$ follows.
Then using Eq.~\eqref{e26860a}, Eq.~\eqref{e26860b}, and Eq.~\eqref{e9323} it 
 is now straightforward to verify that $Z$ satisfies the conditions in 
 Eq.~\eqref{e16127b} of Theorem~\ref{t18481}.

It remains to show that $ZE_?= 0$.
First, with $\Pi_\parallel$ the projector onto $\supp\gamma_1\cap 
 \supp\gamma_2$, we have $ZE_?\Pi_\parallel= Z\Pi_\parallel= 0$ due to 
 $\Pi_\parallel Q_\mu= 0$.
Analogously with $\Pi_\perp$ the projector onto $\ker\mathcal S$, 
 $ZE_?\Pi_\perp= Z\Pi_\perp= 0$.
Thus we only need to show that $\Lambda_\mu ZE_? \Lambda_\nu= 0$.
But this follows from Eq.~\eqref{e12457} and Eq.~\eqref{e3370}.

%==============================================================================%
\section{Appendix: Construction of Jordan bases}\label{a32712}
In this Appendix an explicit construction of Jordan bases (also sometimes 
 called \emph{canonical bases}) of two subspaces $\mathscr A$ and $\mathscr B$ 
 of $\mathbb C^d$ is given (cf.\ also 
 Ref.~\cite{Stewart:1990,Rudolph:2003PRA,Kleinmann:2007JPA,Bergou:2006PRA}).
Jordan bases are orthonormal bases $(\ket{a_i})$ of $\mathscr A$ and 
 $(\ket{b_j})$ of $\mathscr B$, such that $\bracket{a_i}{b_j}= 0$ for all $i\ne 
 j$ and $\bracket{a_k}{b_k}\ge 0$ for all $k$.
With $S_\mathscr{A}$ we denote the $d\times n_\mathscr{A}$ dimensional matrix 
 where the columns are given by the $n_\mathscr{A}$ basis vectors of some 
 orthonormal basis of $\mathscr A$.
Analogously we define $S_\mathscr{B}$ and $n_\mathscr{B}$ by using $\mathscr 
 B$.

Consider a singular value decomposition of
\begin{equation}
S_\mathscr{A}^\dag S_\mathscr{B} = U_\mathscr{A} D U_\mathscr{B}^\dag,
 \mx{i.e.,}
\left(S_\mathscr{A}U_\mathscr{A}\right)^\dag 
\left(S_\mathscr{B}U_\mathscr{B}\right) = D,
\end{equation}
 where $D$ is a $n_\mathscr{A}\times n_\mathscr{B}$ dimensional diagonal 
 matrix.
$U_\mathscr{A}$ and $U_\mathscr{B}$ are unitary matrices.
Let us denote the $i$-th column of $S_\mathscr{A}U_\mathscr{A}$ by $\ket{a_i}$ 
 and the $j$-th column of $S_\mathscr{B}U_\mathscr{B}$ by $\ket{b_j}$.
Then $(\ket{a_i})$ and $(\ket{b_j})$ are Jordan bases of $\mathscr A$ and 
 $\mathscr B$.

\bibliography{the}
\end{document}